\documentclass[prb,twocolumn,eqsecnum,showpacs,preprintnumbers,amsmath,amssymb]{revtex4}

\usepackage{epsfig}
\usepackage{graphicx}
\usepackage{dcolumn}
\usepackage{bm}

\newcommand{\vf}{v_{\rm F}}

\begin{document}

\title{Electronic Properties of Disordered Two-Dimensional Carbon}
\author{N. M. R. Peres$^{1,2}$,
F. Guinea$^{1,3}$, and A. H. Castro Neto$^{1}$}

\affiliation{$^1$Department of Physics, Boston University, 590 
Commonwealth Avenue,
Boston, MA 02215,USA}
\affiliation{$^2$Center of Physics and Departamento de
F\'{\i}sica, Universidade do Minho, P-4710-057, Braga, Portugal}
\affiliation{$^3$Instituto de Ciencia de Materiales de
  Madrid. CSIC. Cantoblanco. E-28049 Madrid, Spain}

\begin{abstract}
Two-dimensional carbon, or graphene, is a semi-metal that 
presents unusual low-energy electronic excitations described in terms
of Dirac fermions.
We analyze in a self-consistent way the effects of localized 
(impurities or vacancies) and extended (edges or grain boundaries) defects on the electronic and
transport properties of graphene. On the one hand, point defects induce a
finite elastic lifetime at low energies with the enhancement of the
electronic density of states close to the Fermi level. Localized disorder
leads to a universal, disorder independent, electrical conductivity at low 
temperatures, of the order of the quantum of conductance. The static 
conductivity increases with temperature and shows oscillations in the 
presence of a magnetic field. The graphene magnetic susceptibility is temperature dependent (unlike an ordinary metal) and also increases with the amount of defects. Optical transport properties are also 
calculated in detail. On the other hand, extended defects induce localized 
states near the Fermi level. In the absence of electron-hole symmetry,
these states lead to a transfer of charge between the defects and the bulk, 
the phenomenon we call self-doping. The role of electron-electron
interactions in controlling self-doping is also analyzed. 
We also discuss the integer and fractional quantum Hall effect in
graphene, the role played by the edge states induced by a magnetic field, 
and their relation to the almost field independent surface states induced at 
boundaries. The possibility of magnetism in graphene, in the presence of
short-range electron-electron interactions and disorder is also analyzed.

\end{abstract}
\pacs{81.05.Uw, 71.55.-i,71.10.-w}

\maketitle

\section{Introduction}          

 Carbon is a life sustaining element that, due to the versatility of 
its bonding, is present in nature in many allotropic forms. Besides being 
an element that is fundamental for life on the planet, it has been explored 
recently for basic science and technology in the form of three-dimensional
graphite, \cite{BCP88} one-dimensional nanotubes, \cite{Retal03} 
zero-dimensional fullerenes, \cite{Setal02} and more recently in 
the form of two-dimensional Carbon, also known as graphene.
Experiments in graphene-based devices have shown that it is possible 
to control their electrical properties
by the application of external gate voltage, 
\cite{Netal04,Betal04,Netal05,Netal05b,Zetal05,Betal05,Zetal05c,Zetal05b}
opening doors for carbon-based nano-electronics. In addition, the
interplay between disorder and magnetic field effects 
leads to an unusual quantum Hall effect predicted theoretically
\cite{PGN05,NGP05,GS05} and measured experimentally 
\cite{Eetal03,Netal05,Zetal05}. 
These systems can be switched from n-type to p-type carriers and 
show entirely new electronic properties. 
We show that their physical properties can be ascribed to 
their low dimensionality, and the phenomenon of self-doping, 
that is, the change in the bulk electronic density due to the breaking
of particle-hole symmetry, and the unavoidable presence of structural defects. Our theory not only provides a description of
the recent experimental data, but also makes new predictions that
can be checked experimentally. 
Our results have also direct implication in the physics of Carbon based
materials such as graphite, fullerenes, and carbon nanotubes.

Graphene is the building block for many forms of Carbon allotropes.
Its structure consists of a Carbon honeycomb lattice made out of hexagons 
(see Fig.~\ref{honey}).
The hexagons can be thought of Benzene rings from which the Hydrogen
atoms were extracted. Graphite is obtained by the stacking of graphene 
layers that is stabilized by weak van der Waals interactions. \cite{P72} 
Carbon nanotubes are synthesized by graphene wrapping. Depending
on the direction in which graphene is wrapped, one can obtain either
metallic or semiconducting electrical properties. 
Fullerenes can also be obtained from 
graphene by modifying the hexagons into pentagons and heptagons in a 
systematic 
way.  Even diamond can be obtained from graphene under extreme pressure
and temperatures by transforming the two-dimensional sp$^2$ bonds into 
three-dimensional 
sp$^3$ ones. Therefore, there has been enormous interest over the years 
in understanding the physical properties of graphene in detail. Nevertheless, 
only recently, with the advances in material growth and control, that one
has been able to study truly two-dimensional Carbon physics.

One of the most striking features of the electronic structure of perfect 
graphene 
planes is the linear relationship between the electronic energy, 
$E_{{\bm k}}$, with the 
two-dimensional momentum, ${\bm k} =(k_x,k_y)$, that is: 
$E_{{\bm k}} =   \vf |{\bm k}|$,
where $\vf$ is the Dirac-Fermi velocity. This singular dispersion relation is a
direct consequence of the honeycomb lattice structure that can be seen
as two interpenetrating triangular sub-lattices. 
In ordinary metals and semiconductors
the electronic energy and momentum are related quadratically via the so-called 
effective mass, $m^*$, ($E_{{\bm k}} =  \hbar^2 {\bm k}^2/(2 m^*)$), 
that controls 
much of their physical properties. Because of the linear dispersion relation,
the effective mass in graphene is zero, leading to a unusual electrodynamics.
In fact, graphene can be described mathematically 
by the two-dimensional Dirac equation, whose elementary excitations are 
particles and holes (or anti-particles), in close analogy with systems 
in particle 
physics. In a perfect graphene sheet the chemical potential, $\mu$, crosses the Dirac point
and, because of the dimensionality, the electronic density of states 
vanishes at the Fermi energy. The vanishing of the effective mass or density of
states has profound consequences.  It has been shown, for instance, that 
the Coulomb 
interaction, unlike in an ordinary metal, remains unscreened \cite{mele} and gives 
rise to an 
inverse quasi-particle lifetime  that increases linearly with energy or 
temperature 
\cite{GGV96}, in contrast with the usual metallic Fermi liquid paradigm, 
where the 
inverse lifetime increases quadratically with energy. 

The fact that graphene is a two-dimensional system has also serious
consequences in terms of the positional order of the Carbon atoms.
Long-range Carbon order in graphene is only really possible at zero temperature
because thermal fluctuations can destroy long-range order in 
two-dimensions 
(the so-called, Hohenberg-Mermin-Wagner theorem \cite{MW66}). 
At a finite temperature $T$, 
topological defects such as dislocations are always present. 
Furthermore, because of the particular structure of the honeycomb lattice,
the dynamics of lattice defects in graphene planes belong to
the generic class of kinetically constrained models\cite{DS00,RS03}, 
where defects are never completely annealed since their number decreases 
only as 
a logarithmic function of the annealing time \cite{DS00}. Indeed, defects
are ubiquitous in carbon allotropes with sp$^2$ coordination and have
been observed in these systems \cite{Hetal04b}. 
As a consequence of the presence of topological defects,
the electronic properties discussed previously, are significantly modified
leading to qualitatively new physics.  
As we show below, extended defects can lead to the phenomenon of self-doping 
with the formation of electron or hole pockets close to the Dirac
points. We show, however, that the presence of such defects can still lead to 
long electronic mean free paths. We present next 
an analysis of the physical properties of graphene as a
function of the density of defects, at zero and finite temperature, frequency, 
and  magnetic field. The defects analyzed
here, like boundaries (edges), dislocations, vacancies, can be considered
strong distortions of the perfect system. In this respect, our work  
complements the studies of defects
and interactions in systems described by the two-dimensional 
Dirac equation \cite{r1}. 

The role of disorder on the electronic properties of coupled graphene
planes shows also its importance on the unexpected appearance of ferromagnetism
in proton irradiated graphite
\cite{Ketal00,Eetal02,MHM02,Ketal03b,Eetal03b,MP05}. 
In a recent publication, the role of the exchange mechanism on
a disordered graphene plane was addressed \cite{PGN05b}.
It was found that disorder can stabilizes a ferromagnetic phase 
in the presence of long-range Coulomb interactions. On the other hand, the effect of disorder on the density of states
of a single graphene plane amounts to the creation of a finite
density of states at zero energy. Therefore, a certain amount of screening
should be present and the question of whether the interplay of disorder and
short-range Coulomb interaction may stabilize a ferromagnetic ground state
has to be addressed as well. 

Moreover, with the current experimental techniques, it is possible to study
not only a single layer of graphene but also graphene multi-layers 
(bilayers, trilayers, etc). 
Recent experiments provide direct 
evidence that while the high-energy physics of graphene multi-layers (for energies 
above around 100 meV from the Dirac point)
is quite different from that of single
layer graphene, the low-energy physics seems to be universal,
two-dimensional, independent
of the number of layers, and dominated by disorder
 \cite{Betal04,Zetal05,Zetal05b}. Hence, the work described here maybe
fundamental for the understanding of this low-energy behavior. 
There is still an interesting question whether this universal low-energy 
physics survives in bulk graphite. 

In this paper we present a comprehensive and unabridged 
study of the electronic properties
of graphene in the presence of defects (localized and extended), and
electron-electron interaction, as a function of temperature, external
frequency, gate voltage, and magnetic field. 
We study the electronic density of states,
the electron spectral function, the frequency dependent conductivity, the
magneto-transport, and the integer and fractional quantum Hall effect.     
We also discuss the possibility of a magnetic instability of graphene
due to short-range electron-electron interactions and disorder 
(the problem of ferromagnetism in the presence of disorder and {\it long-range}
Coulomb interactions was discussed in a previous publication \cite{PGN05b}). 

The paper is organized as follows: in Sec.~\ref{tmatrix}
a formal solution for the single impurity and
many impurities $T-$matrix calculation is given. Details of
the position averaging procedure are given in  
Sec.~\ref{landau}, in connection with the same problem,
but in a magnetic field. In Sec.~\ref{sdos} the problem 
of Dirac fermions in a disordered honeycomb lattice
is studied within the full Born approximation (FBA) and the 
full self-consistent Born approximation (FSBA)
for the density of states. Using the results of Sec.~\ref{sdos}, the spectral
and transport properties of Dirac fermions are computed
in Sec.~\ref{sspec}. In Sec.~\ref{smag} we address the question
of magnetism and the interplay between short-range electron-electron
interactions and disorder.
The density of states of Dirac fermions in a magnetic field perpendicular
to a graphene plane is studied in Sec.~\ref{landau} and the
magneto-transport properties of this system are computed both
at zero and finite frequencies, using the FSBA. The quantization values
for the integer quantum Hall effect and for Jain's
sequence of the fractional quantum Hall effect are 
discussed.  Finally, Sec. \ref{conclusions}
contains our conclusions.
We have also added appendices with the details of the
calculations.

\section{Impurities and vacancies.}
        
\begin{figure}[ht]
\begin{center}
\includegraphics*[width=6cm]{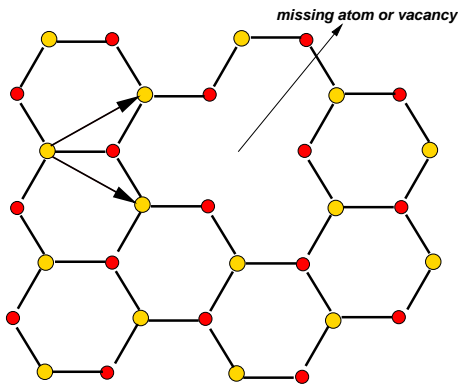}
\end{center}
\caption{\label{honey} (color on line) 
A honeycomb lattice with vacancies, showing its
primitive vectors.}
\end{figure}

The honeycomb lattice can be described in terms of 
two triangular sub-lattices, $A$ and $B$ 
(see Fig.~\ref{honey}).
The unit vectors of the underlying triangular sub-lattice are 
\begin{eqnarray}
\bm a_1 &=& \frac a 2 (3,\sqrt 3,0)\,,
\nonumber \\
\bm a_2 &=& \frac a 2 (3,-\sqrt 3,0)\,,
\end{eqnarray}
where $a$ is the lattice spacing (we use units such that $K_B=1= \hbar$).
The reciprocal lattice vectors are:
\begin{equation}
\bm b_1= \frac {2\pi}{3a} (1,\sqrt 3,0)\,,
\hspace{.2cm}
\bm b_2= \frac {2\pi}{3a} (1,-\sqrt 3,0)\,.
\label{recv}
\end{equation}
The vectors connecting any $A$ atom to its nearest
 neighbors are: 
\begin{eqnarray}
\bm \delta_1 &=& \frac a 2 (1,\sqrt 3,0),
\nonumber \\
\bm \delta_2 &=& \frac a 2 (1,-\sqrt 3,0),
\nonumber \\
\bm \delta_3 &=& a (1,0,0)
\label{nnv}
\end{eqnarray}
and the vectors connecting to next-nearest neighbors
are: 
\begin{eqnarray}
{\bf n}_1 &=& - {\bf n}_2 = {\bf a}_1 \, , 
\nonumber
\\ 
{\bf n}_3 &=& - {\bf n}_4 = {\bf a}_2 \, , 
\nonumber
\\
{\bf n}_5 &=& - {\bf n}_6 = \bm a_1-\bm a_2 \, .
\label{nnnv}
\end{eqnarray}

In what follows we use a tight-binding description for the $\pi$-orbitals
of Carbon with a
Hamiltonian given by:
\begin{eqnarray}
H_{{\rm t.b.}} &=& -t \sum_{\langle i,j \rangle ,\sigma} (a^{\dag}_{i,\sigma} b_{j,\sigma}
+ {\rm h.c.}) \nonumber \\
&+& t' \sum_{\langle \langle i,j \rangle \rangle ,\sigma} (a^{\dag}_{i,\sigma} a_{j,\sigma}
+ b^{\dag}_{i,\sigma} b_{j,\sigma} + {\rm h.c.}) \, ,
\label{Htb}
\end{eqnarray} 
where $a^{\dag}_{i,\sigma}$ ($a_{i,\sigma}$) creates (annihilates) and
electron on site ${\bf R}_i$ with spin $\sigma$ 
($\sigma = \uparrow,\downarrow$) on sub-lattice $A$ and $b^{\dag}_{i,\sigma}$
($b_{i,\sigma}$)  creates (annihilates) and
electron on site ${\bf R}_i$ with spin $\sigma$ 
($\sigma = \uparrow,\downarrow$) on sub-lattice $B$. $t$ is the nearest
neighbor ($\langle i, j \rangle$) 
hopping energy ($t \approx 2.7$ eV), and $t'$ is the
next-nearest neighbor ($\langle \langle i,j \rangle \rangle$) 
hopping energy ($t'/t \approx 0.1$). We notice
{\it en passant} that in earlier studies of graphite \cite{M64} it has
been assumed that $t'=0$. This assumption, however, is not warranted since
there is overlap between Carbon $\pi$-orbitals in the same sub-lattice. 
In fact,  we will show that $t'$ plays an important role in graphene 
since it breaks the
particle-hole symmetry and is responsible for various effects observed
experimentally. 

Translational symmetry is broken by the presence of disorder.
Localized defects such as vacancies and impurities are included
in the tight-biding description by the addition of a 
local energy term:  
\begin{equation}
H_{{\rm imp.}} = \sum_{i,\sigma} V_{i} \left(a^\dag_{i,\sigma}a_{i,\sigma} + 
b^\dag_{i+\bm\delta_3,\sigma}b_{i+\bm\delta_3,\sigma} \right)
\, ,
\label{impurities}
\end{equation}
where $V_{i}$ is a random potential at site ${\bf R}_i$.
In momentum space we define: 
\begin{equation}
 a^\dag_{i,\sigma}=\frac {1}{\sqrt {N_A}}
\sum_{\bm k}e^{i\bm k\cdot \bm R_i}
 a^\dag_{\bm k,\sigma}\,,\hspace{.2cm}
 b^\dag_{i,\sigma}=\frac {1}{\sqrt {N_B}}
\sum_{\bm k}e^{i\bm k\cdot \bm R_i}
 b^\dag_{\bm k,\sigma}\,,
\end{equation}
where $N_A=N_B=N$, and the non-interacting Hamiltonian, $H_1 = H_{{\rm
    t.b.}}+H_{{\rm imp.}}$, reads:
\begin{eqnarray}
H_1&=&\sum_{\bm k,\sigma}[\phi(\bm k)a^\dag_{\bm k,\sigma}b_{\bm k,\sigma}+
\phi^\ast(\bm k) b^\dag_{\bm k,\sigma}a_{\bm k,\sigma}]\nonumber\\
&+&\sum_{\bm k,\sigma} \tilde{\phi}(\bm k) 
(a^\dag_{\bm k,\sigma}a_{\bm k,\sigma}+
b^\dag_{\bm k,\sigma}b_{\bm k,\sigma})
\nonumber \\
&+&
\sum_{\bm q,\bm k,\sigma} V_{{\bf q}} 
[a^\dag_{\bm k+\bm q,\sigma}a_{\bm k,\sigma}
+b^\dag_{\bm k+\bm q,\sigma}b_{\bm k,\sigma}] \, ,
\label{h1}
\end{eqnarray}
where 
\begin{eqnarray}
\phi(\bm k) &=& -t\sum_{i=1}^3 e^{i {\bf k} \cdot {\bf \delta}_i} \, ,
\nonumber
\\
{\tilde \phi}(\bm k)&=&t'\sum_{i=1}^6 
e^{i {\bf k} \cdot {\bf n}_i} \, ,
\end{eqnarray}
and  $V_{{\bf q}}$ is the Fourier transform of the random potential
due to impurities. Hamiltonian (\ref{h1}) is the starting point of
our approach.

\subsection{The single impurity problem and the T-matrix approximation}
\label{tmatrix}   

In the single impurity case one can write $V_{\bf q} = V/N$
where $V$ is the strength of the impurity potential. 
In what follows we use standard finite temperature 
Green's function formalism \cite{DS88,BF04}. Because of
the existence of two sub-lattices, the Green's function
can be written as a $2\times 2$ matrix:  
\begin{eqnarray}
\bm G_{\sigma}(\bm k,{\bf p},\tau) = \left(\begin{array}{cc}
G_{AA,\sigma} (\bm k,{\bf p}, \tau) \hspace{0.5cm} & G_{AB,\sigma}(\bm k,{\bf p},\tau) \\
G_{BA,\sigma} (\bm k,{\bf p},\tau) \hspace{0.5cm} & G_{BB,\sigma}(\bm k,{\bf p},\tau)  
\end{array}\right) \, ,
\end{eqnarray}
where
\begin{eqnarray}
G_{AA,\sigma}(\bm k,{\bf p}, \tau) &=& - \langle {\cal T} a_{\bm k,\sigma}(\tau) a_{\bm p,\sigma}^{\dag}(0)
\rangle \, ,
\nonumber
\\
G_{AB,\sigma}(\bm k,{\bf p}, \tau) &=& - \langle {\cal T} a_{\bm k,\sigma}(\tau) b_{\bm p,\sigma}^{\dag}(0)
\rangle \, ,
\nonumber
\\
G_{BA,\sigma}(\bm k,{\bf p},\tau) &=& - \langle {\cal T} b_{\bm k,\sigma}(\tau) a_{\bm p,\sigma}^{\dag}(0)
\rangle \, ,
\nonumber
\\
G_{BB,\sigma}(\bm k,{\bf p},\tau) &=& - \langle {\cal T} b_{\bm k,\sigma}(\tau) b_{\bm p,\sigma}^{\dag}(0)
\rangle \, ,
\label{defg}
\end{eqnarray}
where $\tau$ is the ``imaginary'' time, and ${\cal T}$ is the time ordering
operator. 

For a single impurity the Green's function can be written as 
$\bm G(\bm k,{\bf p},\tau) = \delta_{\bm k,\bm p} \bm G(\bm k,\tau)$,
where \cite{DS88}:
\begin{equation}
\bm G(\bm k,\omega_n)=\bm G^0(\bm k,\omega_n)+\bm G^0(\bm k,\omega_n)
\bm T_{{\rm imp.}} (\omega_n)\bm G^0(\bm k,\omega_n)\,,
\label{g1imp}
\end{equation}
where $\omega_n = 2 \pi T (n +1/2)$ is the fermionic Matsubara frequency, $\bm G^0(\bm k,\omega_n)$
is the propagator of the tight-binding Hamiltonian (\ref{Htb})
and 
\begin{equation}
\bm T_{{\rm imp.}}(\omega_n)=\frac V N [\bm 1- V \bar {\bm G}^0(\omega_n)]^{-1}\,,
\label{t1imp}
\end{equation}
is the single impurity T-matrix, where:
\begin{equation}
\bar {\bm G}^0(\omega_n)= \frac{1}{N} \sum_{\bm k}\bm G^0(\bm k,\omega_n)\,.
\end{equation}

The above result is exact for a single impurity. For a finite but small
density, $n_i=N_i/N$, of impurities, the Green's function equation becomes:
\begin{equation}
\bm G(\bm k,\omega_n)=\bm G^0(\bm k,\omega_n)+\bm G^0(\bm k,\omega_n)
\bm T(\omega_n)\bm G(\bm k,\omega_n)\,,
\label{finiteimp}
\end{equation}
which is valid up to first order in $n_i$, that is, it takes only into
account the multiple scattering of the electrons by a single impurity.
Equation (\ref{finiteimp}) can be solved as:
\begin{equation}
\bm G(\bm k,\omega_n)=[[\bm G^0(\bm k,\omega_n)]^{-1}-\bm T(\omega_n)]^{-1}\,,
\label{gimp}
\end{equation}
where
\begin{equation}
\bm T(\omega_n)= N_i \bm T_{{\rm imp.}}(\omega_n) = 
V n_i [\bm 1-V \bar {\bm G}^0(\omega_n)]^{-1}\,.
\label{timp}
\end{equation}
For vacancies we take $V \to \infty$ and (\ref{timp})
reduces to:
\begin{equation}
\bm T(\omega_n)= - n_i [\bar {\bm G}^0(\omega_n)]^{-1}\,.
\label{tvac}
\end{equation}

It worth stressing that Eqs. (\ref{g1imp}) and (\ref{t1imp})
although similar in form to 
Eqs. (\ref{gimp}) and (\ref{timp})
 have a very different meaning. Whereas the first set applies
to the single impurity problem, the latter set is the consequence 
of an assemble average over the impurity positions 
(see Sec. \ref{landau} for details on the averaging
procedure in the context of Landau levels)
with a re-summation procedure, corresponding to the
FBA \cite {BF04}.

\subsection{The low-energy physics and the electronic density of states}          
\label{sdos}

The results of the previous subsection are entirely general,
in the sense that no approximation for the band structure
was made. Consider, for simplicity, the tight-binding Hamiltonian
(\ref{Htb}) in the case of $t'=0$, that can be written, in momentum space, as:
\begin{eqnarray}
H_{{\rm t.b.}} =  \sum_{{\bf k},\sigma}
[a^{\dag}_{{\bf k},\sigma},b^{\dag}_{{\bf k},\sigma}] 
\cdot \left[
\begin{array}{cc}
0 \hspace{0.5cm} & \phi({\bf k}) \\
\phi^*({\bf k}) \hspace{0.5cm} & 0
\end{array}
\right]
\cdot 
\left[
\begin{array}{c}
a_{{\bf k},\sigma} \\
b_{{\bf k},\sigma} 
\end{array}
\right] \, ,
\end{eqnarray}
which can be diagonalized and produces the spectrum:
\begin{eqnarray}
E_{\pm}({\bf k}) = \pm |\phi({\bf k})| \, ,
\label{spectra0}
\end{eqnarray}
where the plus (minus) sign is related with the upper (lower) band. 
It is easy to show that the spectrum vanishes at the $K$ point in the 
Brillouin zone
with wave-vector, ${\bf Q} = (2 \pi/(3 \sqrt{3} a),
2 \pi/(3 a))$, and other five points in the Brillouin zone related by
symmetry. In fact, 
it is easy to show that:
\begin{eqnarray}
\phi({\bf Q} +{\bf p})
&\simeq& \frac 3 2 ta e^{i\pi/3}(p_y-ip_x)
\nonumber\\
&+&
\frac 3 8 ta^2e^{i\pi/3}(p_x^2-p_y^2 -2 i p_x p_y) \, ,
\label{asimptf}
\\
\frac{\phi({\bf Q}+{\bf p})}{|\phi({\bf Q}+{\bf p})|} &=& 
e^{i \delta({\bf Q}+{\bf p})}
\approx e^{i\pi/3} \frac{(p_y-i p_x)}{|{\bf p}|} \, , 
\label{coher}
\end{eqnarray}
where $\bm p$ ($p \ll Q$) is measured relatively to the 
$K$ point in Brillouin zone and  we have defined $e^{i\delta(\bm k)}=\phi(\bm k)/\vert\phi(\bm k)\vert$, for latter use. 
Using (\ref{asimptf}) in (\ref{spectra0}) we find:
\begin{equation}
E_{\pm}({\bf Q}+{\bf p})
\simeq \pm \frac 3 2 ta \vert \bm p\vert = 
\pm v_F\vert \bm p\vert \, ,
\label{esympt}
\end{equation}
for the electron's dispersion. Eq. (\ref{esympt}) is the
dispersion of a relativistic particle with ``light'' velocity
$v_F = 3 t a/2$, that is, a Dirac fermion. Hence, at 
low energies (energies much lower than the bandwidth),
the effective description of the tight-binding
problem reduces the 6 points in the Brillouin zone to 2 Dirac
cones, each one of them associated with a different sublattice.  
The low energy description is valid as long as the characteristic
momenta (energy) of the excitations is smaller than a cut-off, $k_c$
($D=v_F k_c$), of the order of the inverse lattice spacing. In the
spirit of a Debye model, where one conserves the total number of
states in the Brillouin zone, we choose $k_c$ such that
$\pi k_c^2=(2\pi)^2/A_c$, where $A_c=3\sqrt{3}a^2/2$ is the area of the 
hexagonal unit cell. Hence, eq.(\ref{esympt}) is valid for $p \ll k_c$
and $E \ll D = k_c v_F$.

So far we have discussed the case of $t'=0$. When $t' \neq 0$ the problem
can also be easily diagonalized and one finds that, close to the K point, 
the electron dispersion changes to:
\begin{eqnarray}
E_{\pm}({\bf Q}+{\bf p}) \approx -3 t' \pm v_F |{\bf p}| + \frac{9 t' a^2}{4}
{\bf p}^2 \, ,
\end{eqnarray}
showing that $t'$ does not change the Dirac physics but
introduces an asymmetry between the upper and lower bands, that is, it breaks the particle-hole symmetry. Hence, $t'$ affects only the intermediate to high
energy behavior and preserves the low-energy physics.
For many of the properties discussed in this section $t'$ does not play
an important role and will be dropped. Nevertheless, we will see 
later that in the
presence of extended defects $t'$ plays an important role and has to be
introduced in order to provide a consistent physical picture of graphene. 

For $t'=0$ we find:
\begin{eqnarray}
{ G}_{AA}(\omega_n,{\bm k})&=& \sum_{j=\pm 1}
\frac{1/2}{i\omega_n - j \vert \phi(\bm k)\vert}
\label{gaa}\,,\\
{ G}_{AB}(\omega_n,{\bm k})&=& \sum_{j=\pm 1}
\frac{je^{i\delta(\bm k)}/2}
{i\omega_n -  j \vert \phi(\bm k)\vert}\label{gab}\,,\\
{ G}_{BA}(\omega_n,\bm k)&=& \sum_{j=\pm 1}
\frac{je^{-i\delta(\bm k)}/2}
{i\omega_n -  j \vert \phi(\bm k)\vert}\label{gba}\,,\\
{ G}_{BB}(\omega_n,{\bm k})
&=& { G}_{AA}(\omega_n,\bm k)\,.
\label{gbb}
\end{eqnarray}
The expansion of the energy around the K point simplifies greatly
the expressions in the calculation of the T-matrix, since they
lead to simpler forms to Eqs. (\ref{gaa})-(\ref{gbb}).
For the case
of vacancies, Eq.(\ref{tvac}), it is easy
to see that at low energies the T-matrix reads:
\begin{equation}
\bm T(\omega_n)=  - n_i [\bar G_{AA}^0(\omega_n)]^{-1} \, \bm I\,,
\end{equation}
where $\bm I$ a 2$\times$2 identity matrix, and
\begin{eqnarray}
\bar G_{AA}^0(\omega_n) &=&  \frac{1}{2N} \sum_{j=\pm 1,{\bf k}} 
\frac{1}{i\omega_n - j \vert \phi(\bm k)\vert}
\nonumber
\\
&=& \frac{1}{2 \rho}\sum_{j=\pm 1} \int \frac{d^2 k}{(2 \pi)^2}
\frac{1}{i\omega_n - j v_F k} 
\nonumber
\\
&=& \frac{1}{4 \pi \rho} \int_0^{k_c}
\frac{dk \, k}{i\omega_n - j v_F k}
\nonumber
\\
&=& - \frac{1}{4 \pi \rho v_F^2} \, \, i\omega_n \ln\left(D^2/\omega_n^2\right) \, ,
\end{eqnarray}
where $\rho = S/V$ is the graphene planar density ($S$ is the area
of the graphene layer). After a Wick rotation
($i\omega_n \to \omega + i 0^+$) one finds: 
\begin{equation}
\bar G^0_{AA}(\omega+i0^+)=-F_0(\omega) -i\pi\rho_0(\omega)\,,
\end{equation}
where, 
\begin{eqnarray}
F_0(\omega)&=&\frac{2 \omega}{D^2} \ln\left(\frac{D}{|\omega|}\right)  \, ,
\label{F}
\\
\rho_0(\omega)&=&
\frac{\vert\omega\vert}{D^2} \, ,  
\label{R}
\end{eqnarray}
where we have used that 
$\rho = 1/A_c = k_c^2/(4 \pi)$, and hence $4 \pi \rho v_F^2
= D^2$. In the above equations we always assume $|\omega| \ll D$.
Notice that $\rho_0(\omega)$ is simply the density of states
of two-dimensional Dirac fermions.

\subsubsection{A single vacancy}          

Assuming that a single unit cell has been diluted, we use Eqs. 
 (\ref{g1imp}) and (\ref{t1imp}) to determine the correction
to the Dirac fermion density of states. The actual density of states,
$\rho(\omega)$, is given by:
\begin{eqnarray}
\rho(\omega) &=& -\frac 1 {\pi} {\rm Im} \bar G_{AA}(\omega+i0^+)
\nonumber
\\
&=&\rho_0(\omega) - \frac {2/N}{D^2}\frac{\rho_0(\omega)}
{F_0^2(\omega)+\pi^2\rho_0^2(\omega)}
\nonumber\\
&\approx& \rho_0(\omega) -\frac {2/N}{\vert\omega\vert\log^2(D/|\omega|)}
\hspace{0.5cm}
(\omega\rightarrow  0)\,,
\label{R1imp}
\end{eqnarray}
indicating that the contribution of the vacancy to the density of states
is singular in the low frequency regime. The contribution is
negative because one has exactly one missing state
associated with the vacancy. The electronic wave function
around a single impurity was computed in Ref.[\onlinecite{Petal05}].
The result obtained here is identical to the one obtained in the
dilution problem in Heisenberg antiferromagnets \cite{CCC01,CCC02}.
The reason for this coincidence is easy to understand: the low energy
excitations of an antiferromagnet in the ordered N\'eel phase 
are antiferromagnetic magnons with linear
dispersion relation, that is, relativistic bosons 
with a ``speed of light'' given by the spin-wave velocity. 
Since we have been
discussing a non-interacting problem, the statistics plays
no role, and the effect of disorder is the same for relativistic
bosons or fermions.   

\subsubsection{The full Born approximation (FBA)}          
\begin{figure}[ht]
\begin{center}
\includegraphics*[width=8cm]{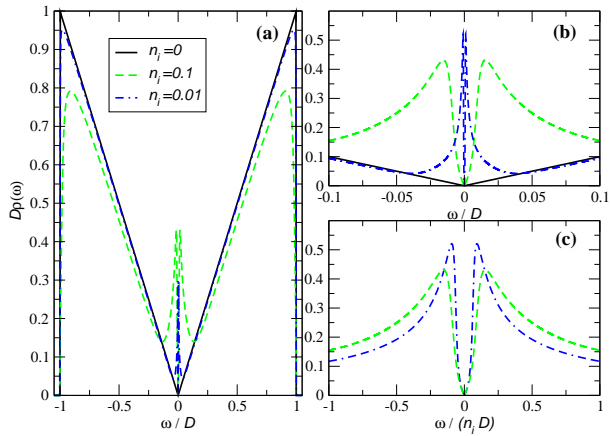}
\end{center}
\caption{\label{fig_RFBA} (color on line) Density of states obtained from
the FBA. Panel {\bf(a)} shows
$\rho(\omega)$ over the entire band;
panel {\bf(b)} shows the low energy
part, where it is seen that the peak in $\rho(\omega)$ 
has a higher value for lower $n_i$; panel   {\bf(c)}
shows that the peak in $\rho(\omega)$ appears
at an energy scale of the order of $n_iD/4$.}
\end{figure}

The situation is clearly different if one has a finite density of
vacancies. In this
case we have to deal with Eqs. (\ref{gimp}) and (\ref{timp})
corresponding to the FBA where all one-impurity
scattering events have been considered. As before, the density of
states is given by 
$\rho(\omega)=-{\rm Im} \bar G_{AA}(\omega+i0^+)/\pi$ and it is possible,
after some tedious algebra,
to obtain an analytical expression for this quantity, given by:
\begin{eqnarray}
\rho(\omega) &=& \frac {\rho_0(\omega)}{D^2}
\frac {2 n_i}{a(\omega)}\ln\left(\frac{D^2}{b^2(\omega)+c^2(\omega)}\right)
\nonumber\\
&+& \frac {1}{\pi D^2}\sum_{\alpha=\pm 1} \frac{\alpha b(\omega)}{a(\omega)}
\left[\arctan\left(\frac{a(\omega)D}{c}\right) \right. 
\nonumber
\\
&+& \left. \arctan\left(\frac{\alpha b(\omega)}{c(\omega)}\right)
\right] \, ,
\label{RFBA}
\end{eqnarray}
with 
\begin{eqnarray}
a(\omega)&=&F_0^2(\omega)+\pi^2\rho_0^2(\omega) \, , 
\nonumber \\
b(\omega)&=& a(\omega)\omega-n_i F_0(\omega) \, ,
\nonumber \\
c(\omega)&=&n_i\pi\rho_0(\omega) \, , 
\label{abcfunc}
\end{eqnarray}
where $F_0(\omega)$ and $\rho_0(\omega)$ are defined in 
(\ref{F}) and (\ref{R}), respectively. A plot of Eq. (\ref{RFBA})
is given in Fig.~{\ref{fig_RFBA}} for two values of the
impurity concentration $n_i$. Once again, the low energy
behavior of $\rho(\omega)$ is the same
found in the context of diluted antiferromagnets.
\cite{CCC01,CCC02}
We remark that the dilution procedure introduces
a low energy scale proportional to $Dn_i$, as can be
seen from panel (c) in Fig.~{\ref{fig_RFBA}}.

\subsubsection{The full self-consistent Born approximation (FSBA)}          

\begin{figure}[hf]
\begin{center}
\includegraphics*[width=8cm]{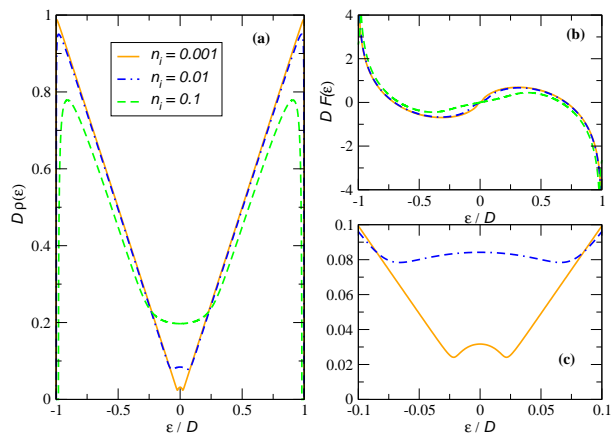}
\end{center}
\caption{\label{fig_RFSCB} (color on line) Density of states obtained in FSBA. Panel
{\bf (a)} shows $\rho(\omega)$; panel {\bf (b)}
shows $F({\omega})$; panel {\bf (c)} shows $\rho(\omega)$ at low energies.
}
\end{figure}

The FBA does not take
into account electronic scattering from multiple
vacancies, it accounts only for multiple scattering
from a single one. In order to include
some contributions from multiple site scattering
another partial series summation can be performed
by replacing the bare propagator in the expression of
the T-matrix in (\ref{tvac}) by full propagator, 
leading to the FSBA. 
Because the matrix elements of the scattering potential
computed from two Bloch states 
$\vert\bm k\rangle$ and $\vert\bm p\rangle$
are assumed momentum independent, the self-energy for the
electrons depends only on the frequency. The self-consistent
problem requires, in general, a careful numerical
solution but in this particular case it is possible
to reduce the problem to a set of coupled algebraic 
equations.
The self-consistent problem requires the solution of
the equation:
\begin{equation}
\Sigma(\omega_n)=\frac {-n_i}{\bar G_{AA}^0(\omega_n-\Sigma(\omega_n))} \, ,
\label{CPA}
\end{equation}
where $\Sigma(\omega_n)$ is the electron self-energy. One can show that
the self-energy can be written as:
\begin{equation}
\Sigma (\omega+i0^+)= \frac {n_i}
{F(\omega)+i\pi\rho(\omega)}\,,
\label{elast}
\end{equation}
where $F(\omega)$ and $\rho(\omega)$ are determined by
the following set of coupled algebraic equations:
\begin{eqnarray}
\label{SCF}
F(\omega)=\frac b{2a(\omega) D^2}\Psi(F,\rho,\omega)+\frac{c(\omega)}{a(\omega)D^2}
\Upsilon(F,\rho,\omega)\, , \\
\label{SCR}
\pi\rho(\omega)=\frac{c(\omega)}{2a(\omega)D^2}\Psi(F,\rho,\omega)-\frac{b(\omega)}{a(\omega)D^2}
\Upsilon(F,\rho,\omega)\, ,
\end{eqnarray} 
where we used the definitions (\ref{abcfunc}) and also defined
the functions $\Psi(F,\rho,\omega)$
and $\Upsilon(F,\rho,\omega)$:
\begin{eqnarray}
\Psi(F,\rho,\omega) \sum_\alpha \ln \left[\frac{(\alpha a(\omega) D+
    b(\omega))^2+c^2(\omega)}{b^2(\omega)+c^2(\omega)} \right] \, ,
\\
\Upsilon(F,\rho,\omega)=-2 \arctan[b(\omega)/c(\omega)] 
\nonumber
\\
+
\sum_\alpha \alpha \arctan[a(\omega)D/c(\omega)-\alpha b(\omega)/c(\omega)]
\, .
\end{eqnarray}
The solution of  Eqs. (\ref{SCF}) and (\ref{SCR})
describes the effect of the vacancies on the density of states
of the Dirac Fermions. $\rho(\omega)$ is the 
self-consistent density of states,
and $F(\omega)$ corresponds to the real part of self-energy
(in analogy with $\rho_0(\omega)$ and $F_0(\omega)$ defined
in (\ref{R}) and (\ref{F})).
In Fig.~\ref{fig_RFSCB} we show
the result of this procedure for various impurity concentrations. 

The low-energy behavior of the density of states, showing a parabolic 
enhancement of $\rho(\omega)$, has also been found
in the context of heavy-fermion superconductors \cite{SMV86}. An exact numerical calculation of the electronic density
was carried out in Ref.[\onlinecite{Petal05}], where it was found that
besides the low energy dome-like shape of the  $\rho(\omega)$ (as shown in
Fig.~\ref{fig_RFSCB}), a large peak appears very close to $\omega=0$.
This peak is reminiscent of the single impurity result given in
(\ref{R1imp}). Hence, besides the peak, the FSBA gives a very good
account of the density of states in this problem.

\begin{figure}[ht]
\begin{center}
\includegraphics*[width=8cm]{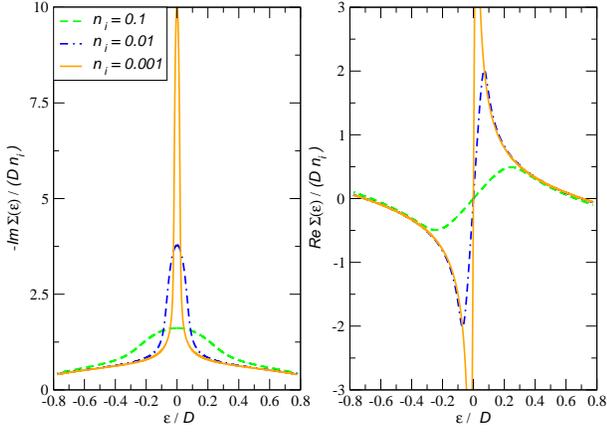}
\end{center}
\caption{\label{fig_self} (color on line) Imaginary (left panel) and real (right panel)
part of the self-energy obtained from
the FSBA.
Note that both quantities are divided by $Dn_i$.}
\end{figure}

Notice that the self-energy, $\Sigma(\omega)$, in (\ref{elast})
depends on $n_i$ in a non-trivial way, since
both the self-consistent $F(\omega)$ and $\rho(\omega)$ also
depend on $n_i$. The self-energy is depicted 
in Fig.~\ref{fig_self} for various values of the dilution 
density $n_i$.

\section{Spectral and transport properties}          
\label{sspec}

The electronic spectral function is defined as:
\begin{eqnarray}
A(\bm k,\omega)=- \frac{1}{\pi} {\rm Im} G(\bm k,\omega+i0^+) \, ,
\end{eqnarray}
and can be interpreted as the probability density that an electron has momentum $\bm k$ and energy $\omega$. For a non-interacting,
non-disordered, problem, the spectral function is simply a Dirac
delta function at $\omega = E({\bf k})$. In the presence of disorder
and/or electron-electron interactions the spectral function is broadened
and its sharpness determines whether the electronic system
supports quasi-particles. The spectral function can be measured directly in angle resolved photoemission experiments (ARPES) \cite{Lanzara05}. 

In terms of the self-energy, $\Sigma({\bf k},\omega)$, the
spectral function reads:
\begin{eqnarray}
A(\bm k,\omega)=- \frac{1}{\pi} \frac{{\rm Im} \Sigma({\bf k},\omega)}{[\omega-E({\bf k})-{\rm Re}\Sigma({\bf k},\omega)]^2+[{\rm Im}\Sigma({\bf k},\omega)]^2} \, .
\end{eqnarray}
In the case of graphene, there are two contributions to the self-energy,
\begin{eqnarray}
\Sigma({\bf k},\omega) = \Sigma_{{\rm e.-e.}} (\bm k) + \Sigma_{{\rm dis.}}(\omega) \, , 
\end{eqnarray} 
where $\Sigma_{{\rm e.-e.}} (\bm k)$
is the self-energy correction due to the electron-electron interactions that was computed originally in Ref.~[\onlinecite{GGV92}]:
\begin{eqnarray}
{\rm Im} \Sigma_{{\rm e.-e.}} (\bm k) =  \frac{1}{48} \left(\frac{e^2}{\epsilon_0 v_F}\right)^2 v_F |\bm k| \, ,
\label{selfee}
\end{eqnarray}
where $e$ is the electron charge, and $\epsilon_0$ the dielectric constant of graphene. The other contribution, $\Sigma_{{\rm dis.}}$,
is due to disorder and is given in (\ref{elast}). 

\begin{figure}[floatfix]
\begin{center}
\includegraphics*[width=8cm]{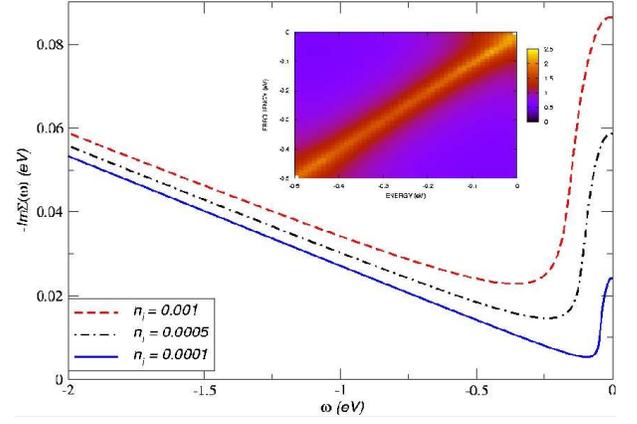}
\end{center}
\caption{\label{fig_spec_func} (color on line) Imaginary part of the electron's
self energy including both the effect of disordered and electron-electron
interaction. The inset shows an intensity plot for the spectral
function $A(\bm k,\omega)$
for $D=8.248$ $eV$, $n_i=0.0001$.}
\end{figure}

Notice that these two contributions to the self-energy have very
different dependence with the energy: while the electron-electron self-energy decreases as the energy (momentum) decreases, the self-energy due to disorder increases as the energy decreases. Hence, electron-electron interactions are dominant at high energies while disorder is dominant at low energies. This interplay between the two self-energies leads to the prediction that there will be a {\it minimum} in the self-energy for
some energy where the electron-electron interaction becomes of the same order of the electron-vacancy interaction. In Fig.~\ref{fig_spec_func} we plot the self-energy as a function of energy for various impurity concentrations together with the spectral function (inset). One can clearly observe the non-monotonic dependence of the self-energy with the energy. This behavior should be
observable in ARPES experiments.

Assuming an electric field applied in the $x$-direction, the frequency dependent (real part) conductivity is calculated from the Kubo formula:
\begin{eqnarray}
\sigma({\bf q},\omega) = \frac{1}{\omega} \int_0^{\infty} dt 
e^{i \omega t} \langle [J^{\dag}_x({\bf q},t),J_x({\bf q},0)] \rangle
\label{kubo}
\end{eqnarray}
where $J_x$ is the $x$-component of the current operator which,
due to gauge invariance, has the form \cite{PK03}:
\begin{equation}
J_x=-ite\sum_{i,\sigma,\bm \delta}\bm u_x\cdot \bm \delta
a^\dag_{i,\sigma}b_{i+\delta,\sigma}
-
\bm u_x\cdot \bm \delta
b^\dag_{i,\sigma+\delta} a_{i,\sigma}
\end{equation} 
(the notation $i+\delta$ means $\bm R_i+\bm \delta$).
In Fourier space, and after expanding the general expression
around the K-point in the Brillouin zone,
we obtain:
\begin{equation}
J_x=-iv_Fe\sum_{\bm k,\sigma}(e^{-i\pi/3}
a^\dag_{\bm k,\sigma}b_{\bm k,\sigma}
-
e^{i\pi/3}
b^\dag_{\bm k,\sigma} a_{\bm k,\sigma})\,.
\label{current}
\end{equation} 
Substitution of (\ref{current}) into (\ref{kubo}) shows that the
problem depends on the Green's functions defined in
(\ref{defg}). However, due to the special
form of Eq. (\ref{coher}) the conductivity
does not have contributions coming from products of Green's 
functions of the form $G_{AB}G_{BA}$. Taking into account
the number of bands and the spin degeneracy, the Kubo formula
for the real part of the conductivity
at finite frequency and temperature 
has the form:
\begin{eqnarray}
\sigma(\omega,T)=- \frac{4 v_F^2e^2}{N A_c \omega}
\int_{-\infty}^{\infty}
\frac {d\epsilon}{2\pi}[f(\epsilon+ \omega)-f(\epsilon)]\times
\nonumber\\
\sum_{\bm k}
{\rm Im} G_{AA}(\bm k,\epsilon+i0^+)
{\rm Im} G_{AA}(\bm k,\epsilon+ \omega+i0^+)\,,
\label{kubofinal}
\end{eqnarray}
where $f(\epsilon) = 1/(e^{(\epsilon-\mu)/T}+1)$ is the
Fermi-Dirac distribution function. The integral over $\bm k$ in (\ref{kubofinal}) can be performed
and find:
\begin{eqnarray}
\sigma(\omega,T)&=&-\frac{e^2}{2 \pi^2 \omega}
\int_{-\infty}^{\infty}
d\epsilon[f(\epsilon+ \omega)-f(\epsilon)] K(\omega,\epsilon) \, ,
\label{sigmat}
\end{eqnarray}
where 
\begin{eqnarray}
K(\omega,\epsilon) = {\rm Im} \Sigma(\epsilon+ \omega) \, \, {\rm Im} \Sigma(\epsilon) \, \, \Theta( \omega,\epsilon)\,,
\label{kern}
\end{eqnarray}
($\Theta(\omega,\epsilon)$ defined in Appendix \ref{ap1}).

It is instructive to consider the zero-temperature, zero-frequency limit
of the conductivity in Eq. (\ref{kubofinal}) (restoring $\hbar$):
\begin{equation}
\sigma_0 = \frac 2{\pi}\frac{e^2}{h}\left (1-
\frac {[{\rm Im} \Sigma(0)]^2}{D^2+[{\rm Im} \Sigma(0)]^2}
\right) \approx \frac 2{\pi}\frac{e^2}{h} \, .
\label{universal}
\end{equation}
The result (\ref{universal}) shows that as long as 
${\rm Im} \Sigma(0)\ll D$, $\sigma_0$ has a universal value
independent of the dilution concentration,
in agreement with earlier theoretical works \cite{F86,L93}, and
in agreement with the experimental data in graphene \cite{Netal05}.

\begin{figure}[floatfix]
\begin{center}
\includegraphics*[width=8cm]{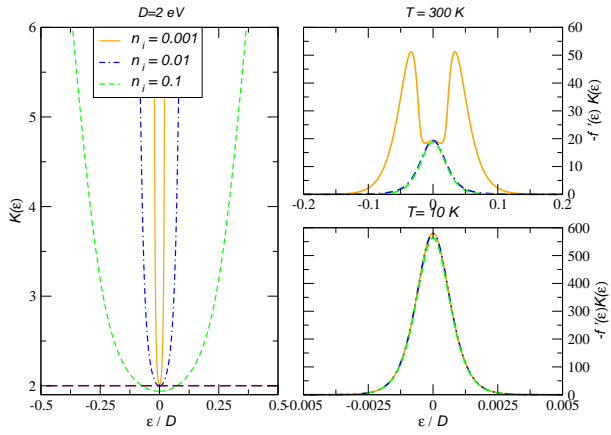}
\end{center}
\caption{\label{fig_kernel_cond} (color on line) Left: Kernel of the integral
for $\sigma(0,T)$ as function of the energy.
Product of the Fermi function derivative
by $K(\epsilon)$ at two different temperatures.}
\end{figure}

\begin{figure}[floatfix]
\begin{center}
\includegraphics*[width=8cm]{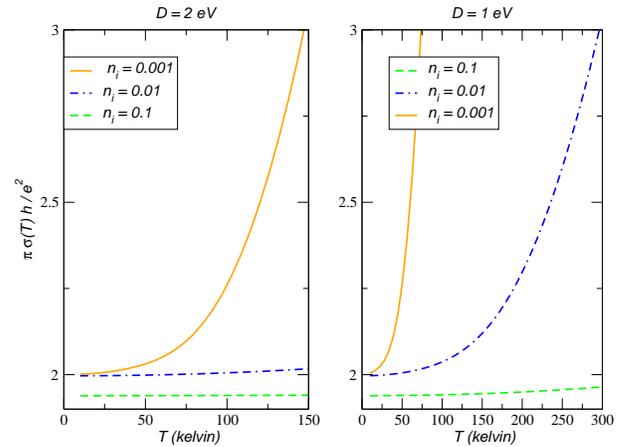}
\end{center}
\caption{\label{fig_cond_T} (color on line) Dependence 
of $\sigma(T,0)$ on the
temperature and on the impurity
dilution $n_i$.}
\end{figure}

At finite temperatures the integral in (\ref{sigmat}) has to
be evaluated numerically. Consider  
$\sigma(0,T)$ whose behavior is determined by
$K(\epsilon) \equiv K(0,\epsilon)$. The quantities $K(\epsilon)$ and $-f'(\epsilon)K(\epsilon)$ (
$-f'(\epsilon)$ is the derivative of the Fermi function 
in order to $\epsilon$) are both represented in Fig.~\ref{fig_kernel_cond}. 
The behavior of  $K(\epsilon)$ shows,  ``V''-like
shape as the energy  $\epsilon$ is varied.
As a consequence, $\sigma(0,0)$ should present
the same  ``V''-like
shape as the chemical potential   $\mu$ moves 
around $\mu=0$.
Such behavior has indeed been observed in
atomically thin carbon films \cite{Netal04,Netal05},
where the density of electrons was controled by a
gate potential.
The temperature dependence of the $\sigma(0,T)$,
for $\mu=0$,
is depicted in Fig.~\ref{fig_cond_T} for different
 vacancy concentrations, and it is found to follow 
Sommerfeld asymptotic expansion, but  the number
of terms needed to fit the numerical curve grows very fast as the dilution
is reduced. 

\begin{figure}[h]
\begin{center}
\includegraphics*[width=8cm]{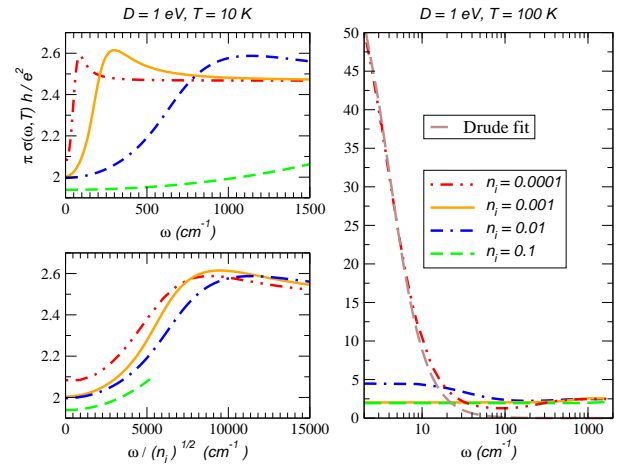}
\end{center}
\caption{\label{fig_cond_omega} (color on line) Dependence 
of $\sigma(T,\omega)$ (in units of $e^2/(\pi \hbar)$) , on the frequency $\omega$.}
\end{figure}

In Fig.~\ref{fig_cond_omega} we plot the frequency dependence
of $\sigma(\omega,T)$ obtained from numerical integration of 
(\ref{sigmat}) with the self-energy given in (\ref{elast}).
At low temperature, we see that  $\sigma(\omega,T)$
develops a maximum around 
an energy value that is dependent on the number of impurities. In fact, if plot $\sigma(\omega,T)$ 
as function of $\omega / \sqrt{n_i}$, the conductivity
almost shows scaling behavior 
for all impurity dilutions (see lower left panel).
As the temperature increases, and if 
$\sigma (0,T)$  is sufficiently large, the conductivity
$\sigma(\omega,T)$ acquires a Drude-like behavior (right panel).

\section{Magnetic response and the role of short range Coulomb interactions}          
\label{smag}

The ferromagnetism measured in proton irradiated graphite
opens the question whether the interplay of interactions
and disorder can drive the system from a paramagnetic 
to a magnetic ground state \cite{Eetal03}. We study the effect of
disorder on the magnetic susceptibility in the presence of short-range interactions, 
and the resulting change in the
tendency towards magnetic instabilities. 
The problem of magnetic instabilities due to long-range exchange
interactions \cite{B29} in the presence of small density of carriers was discussed in 
great detail in ref.~[\onlinecite{PGN05b}]. 
We do not address here the effects
associated to the interplay between the long-range Coulomb interaction and different 
types of lattice disorder\cite{SGV05} and the appearance of 
local moments close to defects \cite{OS91,H01,Letal04,Vetal05}. 

The paramagnetic susceptibility of graphene is given by:
\begin{eqnarray}
\chi (T)&=& \frac {\partial m_z}{\partial h}=4\frac{\partial }{\partial h}
\frac 1 N \sum_{\bm k}\sum_nG_{AA}(\bm k,i\omega_n-h)\nonumber\\
&=&-4\int_{-\infty}^\infty d\epsilon f'(\epsilon)\rho(\epsilon)\,, 
\end{eqnarray}
where 
\begin{equation}
m_z(T)=2\sum_{i,\sigma}\sigma\langle
a^\dag_{i,\sigma}a_{i,\sigma}
\rangle\,,
\end{equation}
is the magnetization, and 
$\rho(\epsilon) $ is the electronic density of states which,
in the presence of disorder, is given in (\ref{SCR}). Within
the Stoner mechanism \cite{S47},
ferromagnetism is possible if the local electron-electron interaction term (the so-called Hubbard term)\cite{Tetal98}, $U$, is large than a critical value given by: 
\begin{eqnarray}
\frac{1}{U_F^c}= \frac{1}{4} \chi(0) \, .
\label{ufc}
\end{eqnarray}
 In the case of an antiferromagnetic
instability the same criteria would lead to another critical value
of $U$ given by:
 \begin{equation}
\frac{1}{U_{AF}^c}= \frac 2{\pi D}
\int_{-\infty}^D d\epsilon \rho(\epsilon)
\arctan \left[\frac {D}{
{\rm Im} \Sigma(\epsilon )} \right]\,.
\label{uafc}
\end{equation}
Notice that in the case of antiferromagnetism one finds that
$U_{AF}^c \approx D/(1-n_i)$ when ${\rm Im}\Sigma \to 0$, in agreement with Hartree-Fock calculations \cite{PAD04}.

\begin{figure}[ht]
\begin{center}
\includegraphics*[width=7cm]{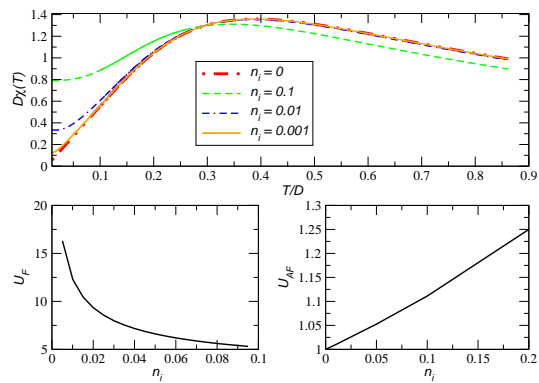}
\end{center}
\caption{\label{fig_sus} (color on line) Top: Dependence 
of $\chi(T)$ on the
temperature  and on the impurity
dilution $n_i$. Bottom: Dependence of $U_F$ and $U_{AF}$
(in units of $D$) as function of $n_i$.}
\end{figure}

In Fig.~\ref{fig_sus} we plot the magnetic susceptibility as function
of $T$ for different values of $n_i$. The signature of the presence
of Dirac fermions comes from the linear dependence on $T$
for $T/D \ll 1$. Notice that, unlike the case of an ordinary metal
that has a temperature dependent Pauli susceptibility, the graphene susceptibility increases with temperatures and number of impurities.
At low temperatures $\chi(T)$ presents a small upturn not visible
in Fig. \ref{fig_sus}.
From the value of $\chi(0)$ and (\ref{ufc}) we obtain the
critical interaction required for a ferromagnetic transition,
which is shown in the lower left panel of Fig.~\ref{fig_sus}. 
Notice that the critical interaction strength for ferromagnetism
decreases as the vacancy concentration increases indicating
that disorder favors a ferromagnetic transition.

Using (\ref{uafc}) and the results of the previous sections we
can also calculate the critical value for an antiferromagnetic
transition. The result is shown in the lower right panel 
of Fig.~\ref{fig_sus}. In contrast with the ferromagnetic case,
the antiferromagnetic instability is suppressed by disorder,
requiring a large value of the electron-electron interaction. 
Notice that the value of the critical ferromagnetic
coupling is always bigger than the antiferromagnetic one,
indicating that at half-filling the graphene lattice is
more susceptible to antiferromagnetic correlations. 
This result is consistent with an old proposal by Linus Pauling
that graphene should be a resonant valence bond (RVB)
state with local singlet correlations \cite{P72}. 

Hence, the Stoner criteria seems to be unable to explain
the ferromagnetic behavior observed experimentally.
One might ask whether additional scattering mechanisms,
such that provided by long-range electron-electron interactions,
can modify the critical values of the couplings. The
self-energy correction due to long-range electron-electron
scattering is given in (\ref{selfee}) and can be added to the
Dyson equation for the Green's function and a new 
self-consistent density of states can be computed. 
This approach does not modify the value of $U_F^c$
which is determined by the low frequency behavior of
the self-energy. In case of antiferromagnetism we find that
indeed it leads to an increase on $U_{AF}^c$, but the result 
is non-conservative since the integral over the
density of states gives a smaller value than $(1-n_i)$.
Therefore, we find from these calculations and previous ones 
\cite{PGN05b} that graphene is not particularly susceptible
to ferromagnetism.

\section{Magneto-Transport}
\label{landau}

The description of the magneto-transport properties of electrons
in a disordered honeycomb lattice is complex because of
the interference effects associated with the Hofstadter problem \cite{GF97}. As in the previous sections,  
we simplify our problem by describing the electrons in the 
honeycomb lattice as Dirac fermions in the continuum.
A similar approach was considered by Abrikosov
in the quantum magnetoresistance study of non-stoichiometric
chalcogenides \cite{A98}.
In the case of graphene, the effective Hamiltonian describing Dirac fermions
in a magnetic field (including disorder) can be written as:
$H=H_0+H_i$ with
\begin{equation}
H_0=-v_F\sum_{i=x,y}\sigma_i[-i \partial_i+eA_i(\bm r)]\,,
\label{llH0}
\end{equation}
where, in the Landau gauge, $(A_x,A_y,A_z)=(-By,0,0)$ is the
vector potential for a constant magnetic field $B$ in the $z-$direction, 
$\sigma_i$ is the $i=x,y,z$ Pauli matrix,
and 
\begin{equation}
H_i=V {\sum_{j=1}^{N_i}\delta(\bm r-\bm r_j) \bm I}\,.
\label{llH1}
\end{equation}
The formulation of the problem in second quantization requires
the solution of $H_0$, which is sketched in Appendix \ref{ap2}.
The field operators are defined as (see  Appendix \ref{ap2} for notation;
the spin index is omitted for simplicity):
\begin{eqnarray}
\Psi(\bm r)&=&\sum_{k}\frac {e^{ikx}}{\sqrt L}
\left(
\begin{array}{c}
 0\\
\phi_0(y)
\end{array}
\right)c_{k,-1} \nonumber\\ 
&+&
\sum_{n,k,\alpha}
\frac {e^{ikx}}{\sqrt {2L}}
\left(
\begin{array}{c}
 \phi_{n}(y-kl_B^2)\\
\phi_{n+1}(y-kl_B^2)
\end{array}
\right)c_{k,n,\alpha}\,.
\end{eqnarray}
The sum over $n=0,1,2,\ldots,$ is cut off at an $n_0$ given by $E(1,n_0)=D$.
In this representation $H_0$ becomes diagonal, leading to 
Green's functions of the form (in Matsubara representation):
\begin{equation}
G_0(k,n,\alpha;i\omega)
=\frac 1{i\omega - E(\alpha,n)}\,,
\end{equation}
is effectively $k$-independent, and $
E(\alpha,-1)=0$ is the zero energy Landau level. The part of
Hamiltonian  due to the impurities is written as:
\begin{widetext}
\begin{eqnarray}
H_i& =& \frac V L\sum_{j=1}^{N_i}\sum_{p,k}e^{-ix_j(p-k)}
\left[\phi_0(y_j-pl_B^2)\phi_0(y_j-kl_B^2) c^\dag_{p,-1}c_{k,-1}
+\sum_{n,\alpha}\frac {\alpha}{\sqrt 2}
\phi_0(y_j-pl_B^2)\phi_{n+1}(y_j-kl_B^2) c^\dag_{p,-1}c_{k,n,\alpha}
\right.
\nonumber\\
&+&
\sum_{n,\alpha}\frac {\alpha}{\sqrt 2}
\phi_{n+1}(y_j-pl_B^2)\phi_0(y_j-kl_B^2) c^\dag_{p,n,\alpha} c_{k,-1}
\nonumber\\
&+&
\left.
\sum_{n,m,\alpha,\lambda}
\frac 1 2[\phi_n(y_j-pl_B^2)\phi_m(y_j-kl_B^2) 
+ \alpha\lambda
\phi_{n+1}(y_j-pl_B^2)\phi_{m+1}(y_j-kl_B^2)]
  c^\dag_{p,n,\alpha} c_{k,m,\lambda}
\right]\,.
\label{hint2}
\end{eqnarray}
\end{widetext}
Equation (\ref{hint2}) describes  the elastic scattering of electrons
in Landau levels by the impurities.  It is
worth noting that this type of scattering connects Landau 
levels of negative and positive energy.

\begin{figure}[ht]
\begin{center}
\includegraphics*[width=8cm]{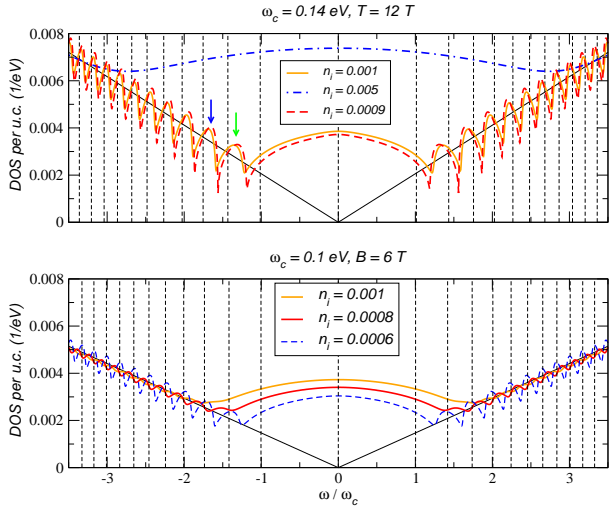}
\end{center}
\caption{\label{fig_dos_landau} (color on line) Electronic density of states in a magnetic field for different dilutions and magnetic field. The 
non-disordered DOS and the position of the Landau levels in the
absence of disorder are also shown. The two arrows in the
upper panel show the position of the renormalized Landau levels
(see Fig.\ref{fig_self_energy}) given by
the solution of Eq. (\ref{rpole}). The energy is given in units
of  $\omega_c\equiv E(1,1)$.}
\end{figure}


\subsection{The full self-consistent Born approximation}          

In order to describe the effect of impurity scattering
on the magnetoresistance of graphene, the Green's function for
Landau levels in the presence of disorder needs to be computed. In the
context of the 2D electron gas, an equivalent study was performed  by Otha and Ando,\cite{O68,O71,Atot}
using the averaging procedure over impurity positions of Duke\cite{D68}.
Here the averaging procedure over impurity positions 
is performed in the 
standard way, namely, having determined the Green's function for a given 
impurity configuration $(\bm r_1,\ldots\bm r_{N_i})$, the position averaged Green's function 
is determined from (as in Sec. \ref{tmatrix}):
\begin{eqnarray}
\langle G(p,n,\alpha;i\omega;\bm r_1,\ldots\bm r_{N_i})\rangle
\equiv G(p,n,\alpha;i\omega) \nonumber\\ 
= L^{-2N_i}\left[\prod_{j=1}^{N_i}
\int d\bm r_j\right] G(p,n,\alpha;i\omega;\bm r_1,\ldots\bm r_{N_i})\,.
\end{eqnarray}
In Sec. \ref{tmatrix} the averaging involved plane wave states; in
the presence of 
Landau levels the average over impurity positions 
involves the wave functions of the one-dimensional 
harmonic oscillator. In the averaging
procedure we have used the following identities:
\begin{eqnarray}
\int dy \phi_n(y-pl_B^2)\phi_m(y-pl_B^2)&=&\delta_{n,m}\,,
\\
\int dp \phi_n(y-pl_B^2)\phi_m(y-pl_B^2)&=&
\frac {\delta_{n,m}}{l^2_B}
\, .
\end{eqnarray}
After a lengthy algebra, the Green's function in the presence of
vacancies, in the FSBA, can be written as:
\begin{eqnarray}
\label{gn}
G(p,n,\alpha;\omega+0^+)&=&[\omega-E(n,\alpha)-\Sigma_1(\omega)]^{-1}\,,
\\
\label{g0}
G(p,-1;\omega+0^+)&=&[\omega-\Sigma_2(\omega)]^{-1}\,,
\label{greenlandau}
\end{eqnarray}
where 
\begin{eqnarray}
\label{S1}
\Sigma_1(\omega)&=&- n_i[Z(\omega)]^{-1}\,,\\
\label{S2}
\Sigma_2(\omega)&=&- n_i[g_cG(p,-1;\omega+0^+)/2 + Z(\omega)]^{-1}\,,\\
\label{Z}
Z(\omega)&=&g_cG(p,-1;\omega+0^+)/2\nonumber\\
&+&g_c\sum_{n,\alpha}G(p,n,\alpha;\omega+0^+)/2\,,
\end{eqnarray}
and $g_c=A_c/(2\pi l_B^2)$ is the degeneracy of a Landau level per
unit cell. One should notice that the Green's functions do not depend
upon $p$ explicitly.
The self-consistent solution of Eqs. (\ref{gn}),
(\ref{g0}), (\ref{S1}), (\ref{S2}) and (\ref{Z}) gives 
density of states, the electron self-energy, and 
the renormalization of Landau level energy position due to disorder.
 
The effect of dilution  in the density of states of Dirac fermions
in a magnetic field is shown in Fig.~\ref{fig_dos_landau}.
For reference we note that $E(1,1)=0.14$ eV, for $B=14$ T,
and $E(1,1)=0.1$ eV, for $B=6$ T.
From Fig.~\ref{fig_dos_landau} we see that given an impurity concentration
the effect of broadening due to vacancies is less effective
as the magnetic field increases. It is also clear that the
position of Landau levels is renormalized relatively to
the non-disordered case. The renormalization of the Landau
level position can be determined from poles of (\ref{greenlandau}):
\begin{equation}
\omega-E(\alpha,n)-{\rm Re}\Sigma(\omega)=0\,.
\label{rpole}
\end{equation}
Of course, due to the importance of scattering at low energies,
the solution to Eq. (\ref{rpole}) does not represent exact
eigenstates of system since the imaginary part of the self-energy
is non-vanishing, however these energies do determine
the form of the density of states, as we discuss below.

\begin{figure}[ht]
\begin{center}
\includegraphics*[width=8cm]{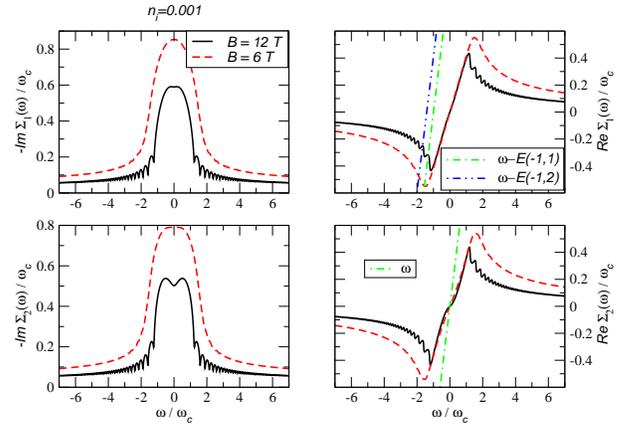}
\end{center}
\caption{\label{fig_self_energy} (color on line) Self-consistent results
for $\Sigma_1(\omega)$ (top) and $\Sigma_2(\omega)$(bottom). 
The energy is given in units of $\omega_c\equiv E(1,1)$. In the left panels we  show the intercept of $\omega-E(\alpha,n)$ with ${\rm Re}\Sigma(\omega)$ as required in (\ref{rpole}).}
\end{figure}

In Fig.~\ref{fig_self_energy}, the graphical solution to
Eq. (\ref{rpole}) is given for two different energies
($E(-1,n)$, with $n=1,2$), being clear that the renormalization 
is important for the first Landau level. This result is due to the
increase of the scattering at low energies, which is present
already in the case of zero magnetic field. The values of $\omega$
satisfying Eq. (\ref{rpole}) show up in density of states
as the energy values where the oscillations due to the Landau level
quantization have a maximum. In Fig.~\ref{fig_dos_landau},
the position of the renormalized Landau levels 
is shown in the upper panel (marked by two arrows), corresponding to the
bare energies $E(-1,n)$, with $n=1,2$. The importance
of this renormalization decreases with the reduction of
number of vacancies. This is clear in  Fig. \ref{fig_dos_landau}
where a visible shift toward low energies is evident
when $n_i$ has a small 10$\%$ change, from $n_i=10^{-3}$ to $n_i=9 \times 10^{-4}$. 


\subsection{Calculation of the transport properties}          
The study of the magnetoresistance properties of the system requires the
calculation of the conductivity tensor. In terms of the field operators,
the current density operator $\bm j$ is given by \cite{GGV96}: 
\begin{equation}
\bm j = v_Fe[\Psi^\dag(x,y)\sigma_x \Psi(x,y),
\Psi^\dag(x,y)\sigma_y \Psi(x,y)]\,,
\end{equation}
leading to current operator in the  Landau basis written as:
\begin{widetext}
\begin{eqnarray}
J_x&=&v_Fe\sum_{p,\alpha}\frac 1 {\sqrt 2}\left[
 c^\dag_{p,-1}c_{k,0,\alpha} + c^\dag_{p,0,\alpha}c_{p,-1}\right]
+v_Fe\sum_{p,n,\alpha,\lambda}
\frac 1 2
\left[
 \lambda (1-\delta_{n,0})c^\dag_{p,n,\alpha} c_{p,n-1,\lambda}
+ \alpha c^\dag_{p,n,\alpha} c_{p,n+1,\lambda}
\right]\,,\\
J_y&=&v_Fe\sum_{p,\alpha}\frac i {\sqrt 2}\left[
 c^\dag_{p,-1}c_{k,0,\alpha} - c^\dag_{p,0,\alpha}c_{p,-1}\right]
+v_Fe\sum_{p,n,\alpha,\lambda}
\frac i 2
\left[
 -\lambda (1-\delta_{n,0})c^\dag_{p,n,\alpha} c_{p,n-1,\lambda}
+ \alpha c^\dag_{p,n,\alpha} c_{p,n+1,\lambda}
\right]\,.
\end{eqnarray}
As in Sec. \ref{sspec}, we compute the current-current correlation function
and from it the conductivity tensor is derived. The diagonal 
component of the conductivity tensor $\sigma_{xx}(\omega,T)$ is given
by (with the spin included):
\begin{eqnarray}
\label{s11}
\sigma_{xx}(\omega,T)&=&-\frac {4(v_Fe)^2}{2\pi  l_B^2}
\frac 1 {\omega}\int_{-\infty}^{\infty}
\frac {d\epsilon}{2\pi}[f(\epsilon+ \omega)-f(\epsilon)]
\left[ \frac 1 2\sum_{\alpha_1}[{\rm Im} G(-1;\epsilon+i0^+)
{\rm Im}  G(0,\alpha;\epsilon+ \omega+i0^+)\right.\nonumber\\
&+&{\rm Im} G(0,\alpha;\epsilon+i0^+)
{\rm Im}  G(-1;\epsilon+ \omega+i0^+)]+
\frac 1 4 \sum_{n\ge 1,\alpha,\lambda}
{\rm Im}  G(n,\alpha;\epsilon+i0^+){\rm Im}  G(n-1,\lambda;\epsilon+ \omega+i0^+)
 \nonumber\\
&+&
\left.
\frac 1 4 \sum_{n\ge 0,\alpha,\lambda}
{\rm Im}  G(n,\alpha;\epsilon+i0^+){\rm Im}  G(n+1,\lambda;\epsilon+ \omega+i0^+)
\right]\,,
\end{eqnarray}
and the off-diagonal component $\sigma_{xy}(\omega,T)$ of the conductivity
tensor is given by:
\begin{eqnarray}
\label{s12}
\sigma_{xy}(\omega,T)&=&\frac {2(v_Fe)^2}{4\pi  l_B^2}
\frac 1 {\omega}\int_{-\infty}^{\infty}
\frac {d\epsilon}{2\pi}\tanh \left(\frac{\epsilon}{2T}
\right)
\sum_{\alpha,\gamma}\left[
\gamma[{\rm Re} G(0,\alpha;\epsilon+\gamma \omega + i0^+){\rm Im} G(-1;\epsilon
+i0^+)\right.\nonumber\\
&-&{\rm Re} G(-1;\epsilon+\gamma \omega + i0^+){\rm Im} G(0,\alpha;\epsilon
+i0^+)]+\sum_{\lambda,n\ge 1}\frac {\gamma}2
[{\rm Re} G(n,\alpha;\epsilon+\gamma \omega + i0^+){\rm Im} G(n-1,\lambda;\epsilon
+i0^+)\nonumber\\
&-&
\left.{\rm Re} G(n-1,\alpha;\epsilon+\gamma \omega + i0^+)
{\rm Im} G(n,\lambda;\epsilon+i0^+)
]\right]\,.
\end{eqnarray}
\end{widetext}
If we neglect the real part of the self-energy, and assume 
 ${\rm Im}\Sigma_{i}(\omega)=\Gamma=$ constant ($i=1,2$),
and let $\omega \to 0$, Eq. (\ref{s11}) reduces to Eq. (85) in Ref. 
[\onlinecite{Getal02}], if we further
assume the case $E(1,1)\gg \Gamma$ then  Eq. (\ref{s12}) reduces
to  Eq. (86) of the same reference.

\begin{figure}[ht]
\begin{center}
\includegraphics*[width=8cm]{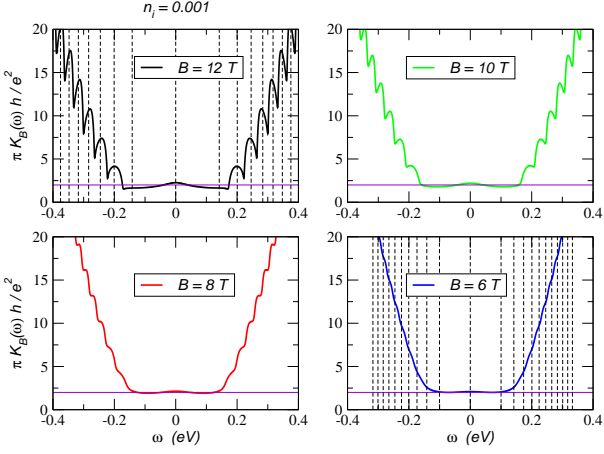}
\end{center}
\caption{\label{fig_kernel_landau} (color on line) Kernel 
of the conductivity 
(in units of $\pi h/e^2$) as function of energy for
different magnetic fields and for a dilution concentration
of $n_i=0.001$. The horizontal line represents the universal limit
$\pi h \sigma_0/e^2=2$.}
\end{figure}

As in Sec. \ref{sspec}, it is instructive to consider
first the case $\omega,T\rightarrow 0$, which leads to
($\sigma_{xx}(0,0)=\sigma_0$):
\begin{eqnarray}
\sigma_0&=&\frac {e^2}h \frac 2{\pi}
\left[\frac {{\rm Im} \Sigma_1(0)/{\rm Im}\Sigma_2(0)-1}{1+({\rm Im}\Sigma_1(0)/\omega_c)^2
}\right.\nonumber\\
&+&\left.\frac {n_0+1}{n_0+1 + ({\rm Im}\Sigma_1(0)/\omega_c)^2}
\right]\,.
\label{s0B}
\end{eqnarray}
When ${\rm Im} \Sigma_1(0)\simeq{\rm Im}\Sigma_2(0)$
and $\omega_c\gg {\rm Im}\Sigma_1(0)$ (or $n_0\gg {\rm Im}\Sigma_1(0)/\omega_c$),
with $\omega_c=E(0,1)=\sqrt 2 v_F /l^2_B$, Eq. (\ref{s0B}) reduces to:
$\sigma_0\simeq 2/\pi (e^2/h)$, which is identical to the result
(\ref{universal}) in the absence of the field found in Sec. \ref{sspec}. This result was obtained previously by Ando and collaborators
using the second order self-consistent Born approximation \cite{ST98,AZS02}.
However, in the FSBA it is required that the above conditions 
be satisfied for this result to hold.  From Fig. \ref{fig_kernel_landau}
we see that the above conditions hold approximately over a
wide ranges of field strength.

Because the d.c. magneto-transport properties have been
measured  graphene samples \cite{Netal04} subjected to a
gate potential (allowing to tune the electronic density), 
it is  important to compute the conductivity kernel,
since this has direct experimental relevance. In the
the case  $\omega\rightarrow 0$ we write the conductivity
$\sigma_{xx}(0,T)$ as:
\begin{equation}
\sigma_{xx}(0,T)= \frac {e^2}{\pi h}\int_{-\infty}^{\infty}
d\epsilon \frac {\partial f(\epsilon)}{\partial \epsilon}
K_B(\epsilon)\,,
\label{sxxb}
\end{equation}
where the conductivity kernel $K_B(\epsilon)$ is given
is Appendix \ref{ap1}. The magnetic field dependence of
kernel $K_B(\epsilon)$
is shown in Fig. \ref{fig_kernel_landau}. Observe 
that the effect of disorder is the creation of a region
where $K_B(\epsilon)$ remains constant before
it starts to increase in energy with superimposed oscillations
coming from the Landau levels. 
The same effect, but with the absence of the oscillations, 
was identified at the
level of the self-consistent density of states plotted
in Fig. \ref{fig_dos_landau}. Together with $\sigma_{xx}(0,T)$,
the Hall conductivity $\sigma_{xy}(0,T)$ allows the calculation
of the resistivity tensor:
\begin{eqnarray}
\rho_{xx} &=& \frac{\sigma_{xx}}{\sigma_{xx}^2+\sigma_{xy}^2} \, ,
\nonumber
\\
\rho_{xy} &=& \frac{\sigma_{xy}}{\sigma_{xx}^2+\sigma_{xy}^2} \, .
\end{eqnarray}

\begin{figure}[ht]
\begin{center}
\includegraphics*[width=8cm]{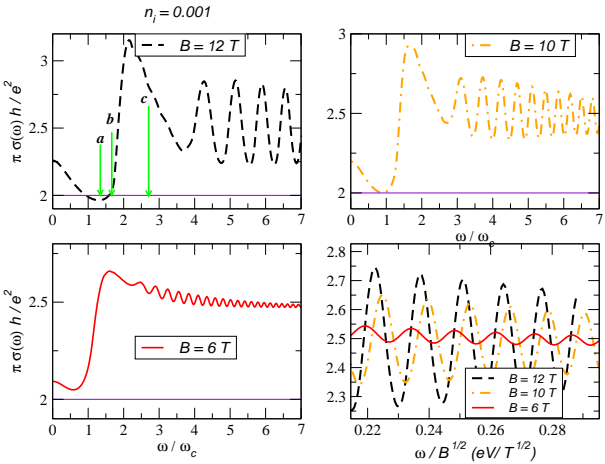}
\end{center}
\caption{\label{fig_sigomega_landau} (color on line) Optical conductivity
(in units of $\pi h/e^2$) at 10 $K$
as function of the energy for
different magnetic fields and for a dilution concentration
of $n_i=0.001$. The vertical arrows in the
upper left panel, labeled {\bf a}, {\bf b},
and {\bf c}, represent the renormalized
$E(1,1)-E(-1,0)$, $E(2,1)-E(-1,0)$, and $E(1,1)-E(1,-1)$
transitions.}
\end{figure}

Let us now focus on the optical conductivity, $\sigma_{xx}(\omega)$.
This quantity can be probed by reflectivity experiments on the sub-THz
to Mid-IR frequency range.\cite{B05}
We depict the behavior of 
Eq. (\ref{s11})   in Fig. \ref{fig_sigomega_landau} for 
different magnetic fields. 
It is clear that the first peak is controlled
by the $E(1,1)-E(1,-1)$, and we have checked it does not
obey any particular scaling form as function of 
$\omega/B$. On the other hand, as the effect of scattering
becomes less important the high energy conductivity 
oscillations 
start obeying the scaling $\omega/\sqrt{B}$, as we show
in the lower right panel of Fig.~\ref{fig_sigomega_landau}.

\section{Extended defects}
\subsection{Self-doping in the absence of electron-hole symmetry.}

The standard description of a graphene sheet, following the usual treatment
of the electronic band structure of graphite\cite{BCP88,SW58,M64,DSM77}, 
assumes that in the absence of interlayer interactions the electronic 
structure of graphite shows electron-hole symmetry. This can be
justified using a tight binding model by considering only hopping between
$\pi$ orbitals located at nearest neighbor Carbon atoms. Within this
approximation it can be shown that in certain graphene edges\cite{WS00,W01}
one would find a flat surface band \cite{Metal05}. Disclinations (a pentagonal or
heptagonal ring) can also lead to a discrete
spectrum and states at zero energy\cite{GGV92,GGV93b}.  
Other types of defects, like a combination of a
five-fold and seven-fold ring (a lattice dislocation) or a Stone-Wales defect 
(made up of two pentagons and two heptagons) also lead to a finite density of 
states at the Fermi level\cite{CEL96,MA01,DSL04}.

Band structure calculations show that the electronic structure of a single
graphene plane is not strictly symmetrical around the energy of the Dirac
points\cite{Retal02}. The absence of electron-hole symmetry shifts the energy
of the states localized near impurities above or below the Fermi level,
leading to a transfer of charge from/to the clean regions to the defects. Hence,
the combination of localized defects and the lack of perfect electron-hole
symmetry around the Dirac points leads to the possibility of self-doping, in
addition to the usual scattering processes whose influence on the transport
properties has been discussed in the preceding sections.

Point defects, like impurities and vacancies, can nucleate a few electronic 
states in their vicinity. Hence, a concentration of $n_i$ impurities 
per Carbon atom leads to a change in the electronic density of the regions 
between the impurities of order
$n_i$. The corresponding shift in the Fermi energy is $\epsilon_{\rm F}
\simeq v_{\rm F} \sqrt{n_{i}}$. In addition, the impurities lead to a
finite elastic mean free path, $l_{\rm elas} \simeq a n_{i}^{-1/2}$, and
to an elastic scattering time $\tau_{\rm elas} \simeq ( v_{\rm F} n_i
)^{-1}$, in agreement with the FSBA calculation in the preceding
sections. Hence, the regions with few impurities can be considered
low-density metals in the dirty limit, as $\tau_{\rm elas}^{-1} \simeq \epsilon_{\rm F}$.

Extended lattice defects, like edges, grain boundaries, or microcracks, are
likely to induce the formation of a number of electronic states proportional
to their length, $L/a$, where $a$ is of the order of the lattice
constant. Hence, a distribution of extended defects of length $L$ at a
distance proportional to $L$ itself gives rise to a concentration of $L/a$
carriers per unit Carbon in regions of order $(L/a)^2$. The resulting system
can be considered a metal with a low density of carriers, $n_{\rm carrier} \propto a/L$ per
unit cell, and an elastic mean free path $l_{\rm elas} \simeq L$. Then, we
obtain:
\begin{eqnarray}
\epsilon_{\rm F} &\simeq &\frac{v_{\rm F}}{\sqrt{a L}} \nonumber \\
\frac{1}{\tau_{{\rm elas}}} &\simeq &\frac{v_{\rm F}}{L}
\end{eqnarray}
and, therefore, $(   \tau_{{\rm elas}} )^{-1} \ll \epsilon_{\rm F}$ when $a/L \ll
1$. Hence, the existence of extended defects leads to the possibility of 
self-doping but maintaining most of the sample in the clean limit. 
In this regime,
coherent oscillations of the transport properties are to be expected,
although the observed electronic properties will correspond to a shifted
Fermi energy with respect to the nominally neutral defect--free system. 

\subsection{Electronic structure near extended defects.}

\begin{figure}[ht]
\begin{center}
\includegraphics*[width=8cm,angle=-90]{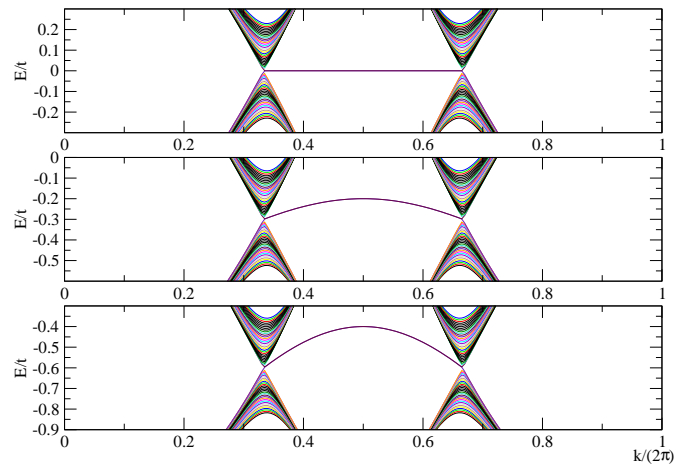}
\end{center}
\caption{\label{ribbon_bands} Bands closest to the Dirac point of a graphene ribbon of 200 unit
  cells width. Top: $t'=0$. Center: $t'=-0.1 t$. Bottom: $t'=-0.2 t$}
\end{figure}

We describe the effects that break electron-hole symmetry near the Dirac
points in terms of a finite next-nearest neighbor hopping between $\pi$
orbitals, $t'$, in (\ref{Htb}). 
From band structure calculations\cite{Retal02}, we expect that $| t'
/ t | \le 0.2$. We calculate the electronic structure of a
ribbon of width $L$ terminated at zigzag edges, which are known to lead to
surface states for $t'=0$. The translational symmetry along the axis of the
ribbon allows us to define bands in terms of the wavevector parallel to this
axis. In Fig.~\ref{ribbon_bands}, we show  the bands closest to $\epsilon=0$
for a ribbon of width 200 unit cells and different values of $t' / t$. The
electronic structure associated to the interior region (the continuum cone),
projected in Fig.~\ref{ribbon_bands} is not significantly changed by
$t'$. The localized surface bands, extending from $k_\parallel = ( 2 \pi )/3$
to $k_\parallel = - ( 2 \pi )/3$, on the other hand, acquires a dispersion of
order $t'$ (for a perturbative treatment of this effect, see ref.~[\onlinecite{SMS05}]). 
Hence, if the Fermi energy remains unchanged at the position of
the Dirac points ($\epsilon_{\rm Dirac} = - 3 t'$), 
this band will be filled, and the ribbon will no longer be
charge neutral. In order to restore charge neutrality, the Fermi level needs
to be shifted down (for the sign of $t'$ chosen in the figure) by an amount
of order $t'$. As a consequence, some of the extended states near the Dirac
points are filled, leading to the phenomenon of self-doping. The local
charge as function of distance to the edges, setting the Fermi energy so that
the ribbon is globally neutral. Note
that the charge transferred to the surface states is very localized near the
edges of the system.

\begin{figure}[ht]
\begin{center}
\includegraphics*[width=7cm,angle=-90]{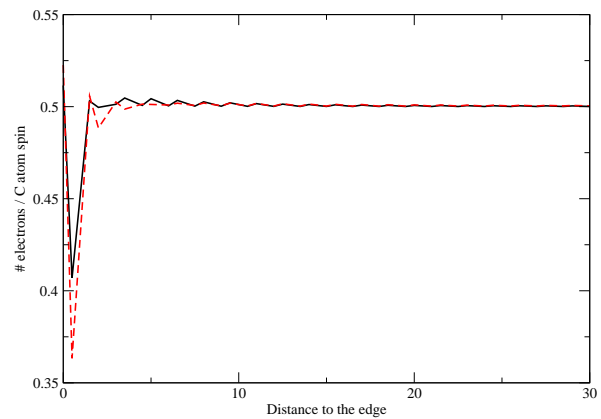}
\end{center}
\caption{\label{ribbon_charge} Charge as function of the distance  to
 the edge (in units of $a$) of a ribbon of width 300 $a$. 
Full line: $t'/t=-0.2$. Broken line: $t'/t=-0.1$. the Fermi
  energy is shifted upwards by $0.054 t$ for $t'=-0.1t$, and $0.077 t$ for $t'=-0.2t$.}
\end{figure}
\begin{figure}[htb]
\includegraphics*[width=7cm,angle=-90]{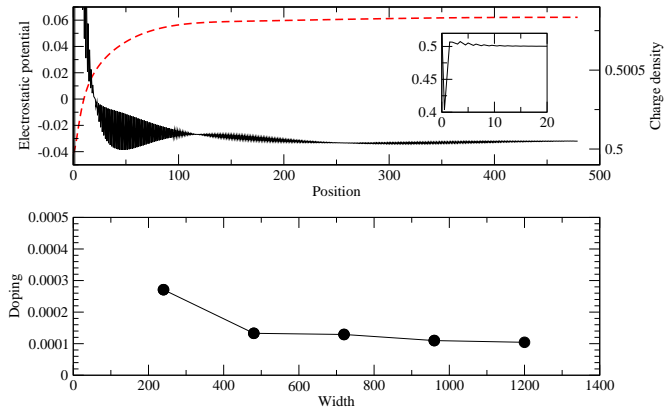}
\caption{\label{dopingfig} Top: Self-consistent charge density (continuous line)
and electrostatic potential (dashed line) of a graphene ribbon with periodic boundary
conditions along the zig-zag edge and with a circumference of size 
$W=80 \sqrt{3} a$, and a length of $L=960 a$.  
 The parameters used are described in the text. The 
inset shows the details of the electronic density near the edge. Due to the presence of the
edge, there is a displaced charge in the bulk (bottom panel) that is
shown as a function of the width $W$.}
\end{figure}

\subsection{Electrostatic effects.}

The charge transfer discussed in the preceding subsection is suppressed by
electrostatic effects, as large deviations from charge neutrality have an
associated energy cost. In order to study these charging effects we add to
the free-electron Hamiltonian (\ref{Htb}) the Coulomb energy of interaction
between electrons:
\begin{eqnarray}
H_I = \sum_{i,j} U_{i,j} n_i n_j \, ,
\label{interact}
\end{eqnarray}
where $n_i = \sum_{\sigma} (a^{\dag}_{i,\sigma} a_{i,\sigma} +
b^{\dag}_{i,\sigma} b_{i,\sigma})$ is the number operator at site
${\bf R}_i$, and
\begin{eqnarray}
U_{i,j} = \frac{e^2}{\epsilon_0 |{\bf R}_i-{\bf R}_j|} \, ,
\end{eqnarray}
is the Coulomb interaction between electrons. 
We expect, on physical grounds, that an
electrostatic potential builds up at the edges, shifting the position of the
surface states, and reducing the charge transferred to/from them. The potential
at the edge induced by a constant doping $\delta$ per Carbon atom 
is roughly, $\sim (\delta e^2/a) (W/a)$ ($\delta e^2/a$ is the Coulomb energy per Carbon),
and  $W$ the width of the ribbon ($W/a$ is the number of Carbons involved). 
The charge transfer is arrested when the potential shifts the localized states to the Fermi energy,
that is, when $t' \approx (e^2/a) (W/a) \delta$. The resulting self-doping is therefore
$\delta \sim ( t' a^2 ) / ( e^2 W )$. 

We treat the Hamiltonian (\ref{interact}) within the Hartree approximation
(that is, we replace $H_I$ by $H_{{\rm M.F.}} = \sum_i V_i n_i$
where $V_i = \sum_j U_{i,j} \langle n_j \rangle$, and solve the
problem self-consistently for $\langle n_i \rangle$).  
Numerical results for graphene ribbons of
length $L = 80 \sqrt{3} a$ and different widths are shown in
 Fig.~\ref{ribbon_charge} and 
Fig.~\ref{dopingfig} ($t'/t= 0.2$ and $e^2/a
= 0.5 t$). The largest width studied is $\sim 0.1 \mu$m, and the total number of
carbon atoms in the ribbon is $\approx 10^5$.  Notice that as $W$ increases, the
self-doping decreases indicating that, for a perfect graphene
plane ($W \rightarrow \infty$), the self-doping effect disappears.
For realistic parameters, we find that the amount of self-doping is $10^{-4} - 10^{-5}$ electrons
per unit cell for domains of sizes $0.1 - 1 \mu$m, in agreement
with the amount of charge observed in these systems.

\begin{figure}[htb]
\includegraphics*[width=7cm,angle=-90]{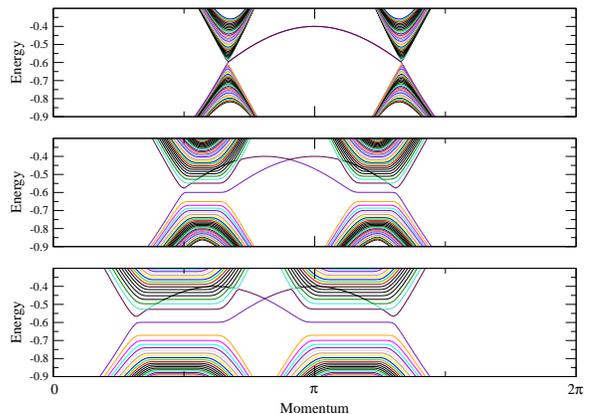}
\caption{\label{boundary_field}  Electronic levels of a graphene ribbon with zig-zag edges in the
  presence of a magnetic field ($t'=-0.2t$).
The magnetic flux per hexagon of the
  honeycomb lattice is $\Phi = 0$ (top), $\Phi= 0.00025$ (center), and $\Phi
  = 0.0005$ (bottom), in units of the quantum of magnetic flux,
  $\Phi_0 = h/e$. The
  corresponding magnetic fields are $0$T, $60$T and $120$T.}
\end{figure}

\subsection{Edge and surface states in the presence of a magnetic field.}  

We can analyze the electronic structure of a graphene ribbon
of finite width in the presence of a magnetic field. The resulting tight
binding equations can be considered as an extension of the Hofstadter problem\cite{H76} to
a honeycomb lattice with edges. The bulk electronic structure is
characterized by the Landau level structure discussed in previous
sections. These states are modified at the edges, leading to chiral edge
states, as discussed in relation to the Integer Hall Effect
(IQHE)\cite{H82}. 
The existence of two Dirac points leads to two independent edge states, 
with the same chirality. In addition, Landau levels with positive energy 
should behave
in an electron-like fashion, moving upwards in energy as their ``center of
gravity'' approaches the edges. Landau levels with negative energy should be
shifted towards lower energy near the edges.

A zig-zag edge induces also a non-chiral surface band. If the width
of these states is much smaller than the magnetic field they will not be much
affected by the presence of the field. The extension of the surface states is
comparable to the lattice spacing for most of the range 
$|k_\parallel| \le ( 2 \pi ) / 3$, except near the Dirac points, so that the
effect of realistic magnetic fields on these states is negligible. 

The finite value of the second nearest neighbor hopping $t'$ modifies the
Landau levels obtained from the analysis of the Dirac equation.
Elementary calculations (as those given in Appendix \ref{ap2})
lead to: 
\begin{equation}
E_{\pm}(n)=-3t'+2l_B^{-2}\alpha\left( n + \frac{1}{2} \right) \pm 
\sqrt{l_B^{-4}\alpha^2+2l_B^{-2}\gamma^2 n}\,,
\label{ELL}
\end{equation}
with $n=0,1,2,3,\ldots$,
$\alpha = 9t'a^2/4$ and $\gamma=3ta/2$, with the single
assumption that  $t\gg t'$. This solution points out a number
of interesting aspects, the most important of which is disappearance
of the zero energy Landau level, made partially of holes and partially
of electrons. With $t'$, the electron or hole nature of the
energy level becomes unambiguous, and half of the
original zero energy Landau level (with $t'=0$)moves down in energy
(relatively to the Fermi energy) and 
the other half moves up. In addition, the level spacing
for electron and hole levels becomes unequal.

\begin{figure}[htb]
\includegraphics*[width=7cm,angle=-90]{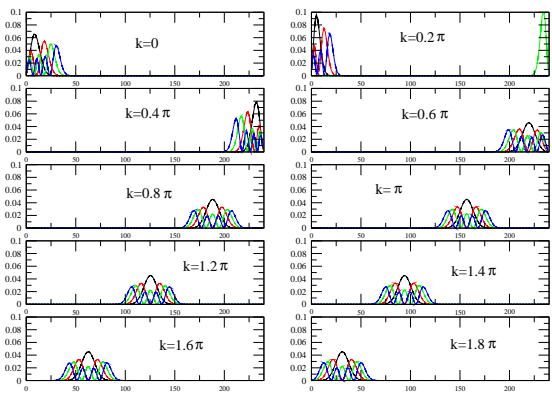}
\caption{\label{boundary_field_charge}  (Color online) Wave function of the lowest Landau levels for
  different momenta.
Black: n=0. Red: n=1. Green: n=2. Blue: n=3.}
\end{figure}

The presence of a magnetic field acting on the ribbon does not break the
translational symmetry along the direction parallel to the ribbon, that
allows us to discuss the electronic structure in terms of the same bands
calculated in the absence of the magnetic field. Results for the ribbon
analyzed in Fig.~\ref{ribbon_bands} for different magnetic fields are shown
in Fig.~\ref{boundary_field}. The ``center of gravity'' of
the wavefunctions associated to the levels moves in the direction transverse
to the ribbon as the momentum is increased. The results show the bulk
Landau levels, and their changes as the wavefunctions approach the edges. The
surface band is practically unchanged, except for small avoided crossings
every time that it becomes degenerate with a bulk Landau level. The results
show quite accurately the expected scaling $\epsilon_n \propto \pm \sqrt{n}$ 
for the eigenenergies derived from the Dirac equation, with small corrections
due to lattice effects and a finite $t'$ (see Appendix
\ref{ap2}). The corresponding wavefunctions for different bands and momenta
are shown in Fig.~\ref{boundary_field_charge}. The Landau levels move
rigidly towards the edges, where one also find surface states.

We can compute the Hall conductivity from the number of chiral states induced
by the field at the edges\cite{H82}. If we fix the chemical potential above
the $n$-th level, there are $2 \times ( 2 n + 1 )$ edge modes crossing the
Fermi level (including the spin degeneracy). Hence, the Hall conductivity is:
\begin{equation}
\sigma_{xy} = \frac{e^2}{h} 2 \times ( 2 n + 1 ) = \frac{4 e^2}{h} (n + 1/2) 
\, .
\label{iqhe}
\end{equation}
This result should be compared with the usual IQHE in heterostructures,
in which the factor of $1/2$ is absent. The presence of this $1/2$ factor is
a direct consequence of the presence of the zero mode in the Dirac fermion
problem. The existence of this anomalous IQHE was predicted long ago
in the context of high energy physics \cite{jackiw84,schakel91} 
and more recently in the context of graphene \cite{PGN05,GS05}, 
but was observed in graphene only recently by two independent groups \cite{Netal05,Zetal05b}.  An incomplete IQHE, with a finite longitudinal resistivity, was observed in HOPG graphite \cite{Ketal03}.

\subsection{Fractional quantum Hall effect}  

While the IQHE depends only of the cyclotron energy, $\omega_c$, 
and therefore is a robust effect, the fractional quantum Hall effect (FQHE)
is a more delicate problem since it is a result of electron-electron
interactions. The problem of electron-electron interactions in the presence
of a large magnetic field in a honeycomb lattice is a complex problem
that deserves a separate study. In this paper we make a few conjectures
about the structure of the FQHE based on generic properties of Laughlin's
wave functions. 

The electrons occupying the lowest Landau level are assumed to be in a
many-body wavefunction written as (${\bf R}_i = (x_i,y_i)$ and $z=x+iy$) \cite{MS03} :
  \begin{equation}
\Psi=\exp(-i 2 m \sum_{i<j}\alpha_{i,j}) 
\Phi(z_1,\ldots,z_N)\,,
\end{equation}
where $\alpha_{i,j}=\arctan [{\rm Im}(z_i
-z_j)/{\rm Re}(z_i-z_j)]$, $\Phi
(z_1,\ldots,z_N)$ is an anti-symmetric function
of the interchange of two $z's$, and 
$m=0,1,2,\ldots$.
The effect of the singular phase associated with the many-body wave function 
is to introduce an effective 
magnetic field $B^\ast$ given by:
\begin{equation}
B^\ast = B - \bm\nabla  \bm a(\bm r)/e=B - 2\pi (2 m) \rho(\bm r)/e\,,
\end{equation}
where the gauge field $ \bm a(\bm r)$ is given by 
\begin{equation}
\bm a(\bm r_i) =  2 m \sum_{j\ne i}\bm\nabla(\bm r_i)\alpha_{i,j}\,,
\end{equation}
and $\rho(\bm r)$ is the electronic density. The procedure outlined
above is called flux-attachment and leads to appearance of composite
fermions. These composite particles do not feel the 
external field $B$ but instead an effective field $B^{\ast}$. 
Therefore, the FQHE of electrons can be seen as an IQHE of these composite particles.
 
Given an electronic density $\delta $, we may define 
an effective filling factor $p^\ast$ for the composite particles as:
\begin{equation}
p^\ast = \frac {2 \pi \delta}{e B^\ast}\,.
\label{pef}
\end{equation}
In the lowest Landau level the electron filling factor is:  
\begin{equation}
p = \frac {2 \pi \delta} {e B}\,,
\label{p}
\end{equation}
and combining Eqs. (\ref{pef}) and (\ref{p}) we obtain: 
\begin{equation}
p = \frac {p^\ast}{2 m p^\ast +1}\, ,
\end{equation}
so, we can write:
\begin{eqnarray}
B^{\ast} = B (1 - 2 m p) \, .
\label{bast}
\end{eqnarray}

The crucial assumption in the case of graphene is that the effective
$p^\ast$ associated with the integer quantum Hall effect of composite
particles has the form given in (\ref{iqhe}), that is (spin ignored):
\begin{equation}
p^\ast =(2n+1)\,,\hspace{1cm} n=0,1,2,3,\ldots \,,
\label{ass}
\end{equation}
(the effective field $B^\ast$ is such that the system has one 
or more filled composite particle Landau levels, and the chemical potential lies between two these) leading to a quantized Hall conductivity given by:
\begin{equation}
\sigma_{xy}= \frac {2n+1}{2 m(2n+1)+1}\frac{2 e^2}{h} \, .
\label{fqheII}
\end{equation}
For $n=0$, one obtains the so-called Laughlin sequence:
$\sigma_{xy}= 1/(2 m + 1) (2 e^2/h)$, and for $m=0$ we
recover (\ref{iqhe}).
This argument shows that Jain's sequence is quite different from that of the 2D electrons gas \cite{J89}. 

As in the case of the IQHE, the FQHE can be thought in terms of
chiral edge states, or chiral Luttinger liquids, 
that circulate at the edge of the sample \cite{wen}.
One can see the IQHE and FQHE as direct consequences of the presence
of these edge states. 
Because of their chiral nature, edge states do not localize in the
presence of disorder and hence the quantization of the Hall conductivity
is robust. In graphene, as we have discussed previously, zig-zag edges
support surface states that are non-chiral Luttinger liquids. We have
recently shown that electron-electron interactions between chiral Luttinger
liquids and non-chiral surface states can lead to instabilities of the
chiral edge modes leading to edge reconstruction \cite{NGP05} and
hence to the destruction of the quantization of conductivity. We also
have shown that this edge reconstruction depends strongly on the 
amount of disorder at the edge of the sample. While this effect is
not strong in the IQHE (because the cyclotron energy is very large
when compared with the other energy scales), it makes the experimental observation of the FQHE in graphene very difficult.


\section{Concluding Remarks}
\label{conclusions}

To summarize, we have analyzed the influence of local 
and extended lattice defects 
in the electronic properties of single graphene layer. 
Our results show that: 
(1) Point defects, such as vacancies, lead to an enhancement
of the density of states at low energies and to a finite density
of states at the Dirac point (in contrast to the clean case where
the density of states vanishes); (2) Vacancies have a strong
effect in the Dirac fermion self-energy leading to a very short
quasi-particle lifetime at low energies; (3) The interplay between
local defects and electron-electron interaction lead to the existence
of a minimum in the imaginary part of the electron self-energy 
(a result that can be measured in ARPES); (4) The low temperature d.c. conductivity is a universal number, independent on the disorder concentration and magnetic field; (5) The d.c. conductivity, as in the case of a semiconductor, increases with temperature and chemical potential (a result that can be observed by applying a bias voltage to the system); (6) The a.c. conductivity increases with frequency at low frequency and at very low impurity concentrations can be fitted by a Drude-like model; (7) The magnetic susceptibility of graphene increases with temperature (it is not Pauli-like, as in an ordinary metal) and is sensitive to the amount of disorder in the system (it increases with disorder); (8) Within the Stoner criteria for magnetic instabilities
we find that graphene is very stable against magnetic ordering and that the phase diagram of the system is dominated by paramagnetism; (9) In the presence of a magnetic field and disorder, the electronic density of states shows oscillations due to the presence of Landau levels which are shifted from their positions because of disorder; 
(10) The magneto-conductivity presents oscillations in the presence of fields and that their dependence with chemical potential and frequency are rather non-trivial, showing transitions between different Landau levels; (11) Extended defects, such as edges, lead to the effect of self-doping where charge is transfered from/to the defects to the bulk in the absence of particle-hole symmetry; (12) The effect of extended defects on transport is very weak and that electron scattering is dominated by local defects such as vacancies; (13) The quantization of the Hall conductance in the IQHE is anomalous relative to the case of the 2D electron gas with an extra factor of $1/2$ due to the presence of a zero mode in the Dirac fermion dispersion; (14) We conjecture that the FQHE in graphene has a sequence of states which is very different from the sequence found in the 2D electron gas and we propose a formula for that sequence. 

The results and experimental predictions made in this work are based on a careful analysis of the problem of Dirac fermions in the presence of disorder, electron-electron interactions and external fields. We use well established theoretical techniques and find results that agree quite well with a series of amazing new experiments in graphene \cite{Netal04,Betal04,Zetal05c,Netal05,Zetal05,Netal05b,Zetal05b}. The main lesson of our work is that graphene presents a completely new electrodynamics when compared to ordinary metals which are described quite well within Landau's Fermi liquid theory. In this work, we focus on the effects of disorder and electron-electron interaction and have shown that Dirac fermion respond to these perturbations in a way which is quite different from ordinary electrons. In fact, graphene is a non-Fermi liquid material where there is no concept of an effective mass and, therefore, a system where Fermi liquid concepts are not directly applicable. A new phenomenology, beyond Fermi liquid theory, has to be developed for this system. Our work can be considered a first step in that direction.


\section*{Acknowledgments}
The authors would like to thank D. Basov, 
W. de Heer, A. Geim, G.-H. Gweon, P. Kim, A. Lanzara, Z.Q. Li, 
J. Nilsson, V. Pereira, J. L. Santos,  
S.-W. Tsai, and S. Zhou for many useful discussions.
N.M.R.P and  F. G. are thankful to the Quantum Condensed Matter
Visitor's Program at Boston University.
A.H.C.N. was supported through NSF grant DMR-0343790.
N. M. R. P. thanks 
Funda\c{c}\~ao para a Ci\^encia e Tecnologia for a sabbatical grant
partially supporting his sabbatical leave,  
the ESF Science Programme INSTANS 2005-2010, 
and 
FCT under the grant POCTI/FIS/58133/2004.


\appendix
\section{$\Theta(\omega,\epsilon)$,
and $K_B(\epsilon)$}
\label{ap1}
In the calculation of $\sigma(\omega,T)$ and
$\sigma(0,T)$ we defined:  
\begin{widetext}
\begin{eqnarray}
\Theta(\omega,\epsilon)
&=&\sum_{s_1=\pm 1,s_2=\pm 1}
\left[
\frac {N+s_1M}B\left(
\arctan\left[\frac {D-s_1A}{B}\right]+\arctan\left[\frac {s_1A}{B}\right]
\right)+
\frac {P+s_2V}E\left(
\arctan\left[\frac {D-s_2C}{E}\right]+\arctan\left[\frac {s_2C}{E}\right]
\right)\right.
\nonumber\\
&+&
\left.
\frac M 2 \log \left[\frac {(D-s_1A)^2+B^2}{A^2+B^2}\right]
+
\frac V 2 \log \left[\frac {(D-s_1C)^2+E^2}{C^2+E^2}\right]
\right]\,,
\end{eqnarray}
\end{widetext}
where
\begin{eqnarray}
M&=&(C^2+E^2-A^2-B^2)/{\cal D},\hspace{0.5cm}V=-M\nonumber\,,\\
N&=&2(s_1A-s_2C)(A^2+B^2)/{\cal D}\nonumber\,,\\
P&=&-2(s_1A-s_2C)(A^2+B^2)/{\cal D}\nonumber\,,\\
{\cal D}&=&(A^2+B^2-C^2-E^2)^2\nonumber\, \\
&+&4(A^2+B^2)(C^2-s_1s_2AC)\nonumber\,,\\
&+&4(C^2+E^2)(A^2-s_1s_2AC)\,,\nonumber\\
A&=&\epsilon+\omega-{\rm Re}\Sigma(\epsilon+\omega)\nonumber\,,\\
B&=&{\rm Im} \Sigma(\epsilon+\omega)\nonumber\,,\\
C&=&\epsilon-{\rm Re}\Sigma(\epsilon)\nonumber\,,\\
E&=&{\rm Im} \Sigma(\epsilon)\,,\nonumber\\
\end{eqnarray}
and  $\cos\alpha=(C^2-E^2)/(C^2+E^2)$.

In the magneto-transport properties, $\sigma_{xx}(0,T)$ given by Eq. (\ref{sxxb}), depends on the kernel
$K_B(x)$, which is defined as:
\begin{widetext}
\begin{eqnarray}
K_B(x)&=&\frac {v_F^2}{2\pi l_B^2}\sum_\alpha\left[
\frac {{\rm Im}\Sigma_2(x)}{[x -{\rm Re}\Sigma_2(x)]^2+[{\rm Im}\Sigma_2(x)]^2}
\frac {{\rm Im}\Sigma_1(x)}{[x - E(\alpha,0)-{\rm Re}\Sigma_1(x)]^2+[{\rm Im}\Sigma_1(x)]^2}
\right.\nonumber\\
&+&
\sum_{\lambda,n\ge 1}\left.
\frac {{\rm Im}\Sigma_1(x)}{[x -E(\alpha,n)-{\rm Re}\Sigma_1(x)]^2+[{\rm Im}\Sigma_1(x)]^2}
\frac {{\rm Im}\Sigma_1(x)}{[x -E(\lambda,n-1)-{\rm Re}\Sigma_1(x)]^2+[{\rm Im}\Sigma_1(x)]^2}
\right] \, .
\end{eqnarray}
\end{widetext}
\section{The Dirac equation in a magnetic field}
\label{ap2}

The Hamiltonian (\ref{llH0}) can be solved using a trial spinor of the form:
\begin{equation}
\psi(\bm r)\left(
\begin{array}{c}
c_1\phi_1(y)\\
c_2\phi_2(y)
\end{array}
\right)\frac {e^{ik_xx}}{\sqrt{L}}\,,
\end{equation}
with $L$ the size of the system in the $x$ direction.
After straightforward manipulations, the eigenproblem reduces to:
\begin{equation}
{v_F \sqrt 2}{l_B}
\left(
\begin{array}{cc}
0 & a \\
a^\dag&0
\end{array}
\right)
\left(
\begin{array}{c}
c_1 \phi_1(y)\\
c_2 \phi_2(y)
\end{array}
\right)
E
\left(
\begin{array}{c}
c_1 \phi_1(y)\\
c_2 \phi_2(y)
\end{array}
\right)\,,
\label{h2}
\end{equation}
where
\begin{eqnarray}
a &=& \frac {1}{\sqrt 2 l_B}(y+l^2_B\partial_y)\,,
\label{a}
\\
a^\dag &=& \frac {1}{\sqrt 2 l_B}(y-l^2_B\partial_y)\,,
\label{ad}
\end{eqnarray}
with the magnetic length defined as $l^2_B=1/(eB)$.
For the case
of $E\ne 0$, it is simple to see that the spinor
\begin{equation}
\frac 1 {\sqrt{2}}\left(
\begin{array}{c}
 \phi_n(y)\\
\alpha \phi_{n+1}(y)
\end{array}
\right)\,,
\end{equation}
is an eigenfunction of (\ref{h2})
with eigenvalue  
$E(\alpha,n)=\alpha v_F \sqrt{2}/l_B\sqrt{n+1}$,
with $\alpha=\pm 1$, and $\phi_n$ ( $n=0,1,2,\ldots$) the $n$
eigenfunction of the usual 1D harmonic oscillator.
In addition, 
there exists a zero energy mode whose eigenfunction is given by:
\begin{equation}
\left(
\begin{array}{c}
 0\\
\phi_0(y)
\end{array}
\right)\,,
\end{equation}
that completes the solution of the original eigenproblem.
As in the more conventional Landau level problem, the degeneracy
of each level is $L^2B/\phi_0$, with $\phi_0=h/e$ the quantum of flux.

\bibliography{graphite0_1}

\end{document}